\documentclass[floatfix,twocolumn,showpacs,preprintnumbers,amsmath,amssymb]{revtex4}

\usepackage{dcolumn}
\usepackage{bm}
\usepackage{graphicx}
\usepackage{times}

\def\hfq{\hfill\quad}
\def\cc#1{\hfq#1\hfq}
\def\tvi{\vrule height 12pt depth 5pt width 0pt}
\def\traithorizontal{\noalign{\hrule}}
\def\tv{\tvi\vrule}

\def\eg{{\it e.g.}\ }  \def\ie{{\it i.e.}\ }
  
 \def\rms{{\it r.m.s.}\ }
\def\viz{{\it viz.}\ }

\newcommand{\be}{\begin{equation}}
\newcommand{\ee}{\end{equation}}

\newcommand{\bA}{\mathbf{A}} \newcommand{\bB}{\mathbf{B}}
\newcommand{\Asz}{A_{s_z}} 
\newcommand{\alp}{\alpha}
\newcommand{\alpmm}{\alpha^{-1}_m} \newcommand{\alpmv}{\alpha^{-1}_k}
 
\newcommand{\bj}{\mathbf{j}} \newcommand{\bk}{\mathbf{k}}
\newcommand{\bom}{\mbox{\boldmath $\omega$}}

 \newcommand{\bv}{\mathbf{v}}
\newcommand{\bAs}{\mathbf{A_s}} \newcommand{\bjs}{\mathbf{j_s}} 
\newcommand{\bus}{\mathbf{u_s}} \newcommand{\bBs}{\mathbf{B_s}} 
\newcommand{\boms}{\mbox{\boldmath $\omega_s$}}

\newcommand{\bx}{\mathbf{x}} \newcommand{\bxp}{\mathbf{x^{\prime}}}

\newcommand{\Rla}{R_{\lambda}}

\def \pmbtext#1{\leavevmode \setbox0\hbox{#1}
     \kern-0,2pt \copy0 \kern-\wd0 \kern0,4pt \copy0 \kern-\wd0
     \kern-0,2pt \raise0,3pt \box0 }

\begin{document}
\title{A numerical study of the alpha model 
for two-dimensional magnetohydrodynamic turbulent flows}

\author{Pablo D. Mininni$^1$, David C. Montgomery$^2$ and 
        Annick G. Pouquet$^1$}

\affiliation{$^1$ Advanced Study Program, National Center for Atmospheric 
Research, P.O. Box 3000, Boulder, Colorado 80307, U.S.A.}
\affiliation{$^2$ Dept. of Physics and Astronomy,
Dartmouth College, Hanover, NH 03755, U.S.A.} 

\begin{abstract}
We explore some consequences of the ``alpha model,'' also called the
``Lagrangian-averaged'' model, for two-dimensional incompressible 
magnetohydrodynamic (MHD) turbulence.
This model is an extension of the smoothing procedure in 
fluid dynamics which filters velocity fields locally while leaving their 
associated vorticities unsmoothed, and has proved useful for high Reynolds 
number turbulence computations. We consider several known effects (selective 
decay, dynamic alignment, inverse cascades, and the probability distribution
functions of fluctuating turbulent quantities) in magnetofluid turbulence
and compare the results of numerical solutions of the primitive MHD equations
with their alpha-model counterparts' performance for the same flows, in regimes
where available resolution is adequate to explore both. The hope is to justify 
the use of the alpha model in regimes that lie outside currently available 
resolution, as will be the case in particular in three-dimensional geometry or
for magnetic Prandtl numbers differing significantly from unity.
We focus our investigation, using direct numerical simulations with
a standard and fully parallelized pseudo-spectral method and periodic boundary 
conditions in two space dimensions, on the role that such a 
modeling of the small scales using the Lagrangian-averaged framework 
plays in the large-scale dynamics of MHD turbulence.
Several flows are examined, and for all of them one can conclude that 
the statistical properties of the large-scale spectra are recovered, whereas 
small-scale detailed phase information (such as \eg the location of 
structures) is lost.
 
\end{abstract}
\maketitle

\section{Introduction}

One of the most persistent difficulties in the computation of the turbulent 
behavior of fluids and magnetofluids is the wide range of dynamically 
interacting length and time scales that have to be evaluated. At large 
Reynolds 
number, many orders of magnitude in length scales are implied, for example, 
in the dynamical behavior of the 
atmosphere, the oceans, or the solar wind, to take some familiar situations. 
For many purposes, it might be adequate to compute only the long-wavelength 
components of the spectra of the fields involved if some more economical 
representation or model of the small scale behavior could be given which would 
not do violence to the accuracy with which the large scales are computed. Such 
topics as ``large eddy simulation'' and ``eddy viscosity,'' designed to cope 
with this difficulty, have generated a vast literature, one which we make 
no attempt to survey here (see \eg Refs. \cite{moin84}-\cite{MK00}).

A novel offering along these lines which has appeared in 
recent years is the so-called ``alpha model'' of Holm, Foias, Margolin, 
Marsden, Olson, Ratiu, Titi, Wynne and especially Chen, whose comparisons 
with turbulent channel and pipe flow called the most attention to the 
alpha model's possibilities (\eg, Refs. \cite{HM98}-\cite{Foias2001}; many 
other references could also be cited).
The model is also variously called the ``Lagrangian averaged model'' 
or, in some of the earliest papers, the ``Camassa-Holm'' equations. 
This alpha model is subject to a variety of 
derivations, interpretations and connections,
ranging from the mathematically sophisticated 
\cite{HM98,H01,AM78} to the intuitive and simple \cite{MP02}.
It is of interest to subject its predictions to tests against both 
experimental data (see Ref. \cite{CFHOTW99}) and numerical solutions of the 
relevant 
continuum equations to which alpha modeling has not yet been applied 
(see \eg Ref. \cite{CHMZ99} for the Navier-Stokes equations in three 
dimensions). Its extension to the case of coupling to a magnetic field in the 
magnetohydrodynamic (MHD) limit and in the non-dissipative case can be 
found in 
\cite{H01,HMHD2}. In that context, the main purpose of this article is to 
carry out some of the numerical tests that have not previously been done
for the case of MHD.

It is to be emphasized that there is no derivation of the alpha 
model that is completely systematic and deductive. Every presentation 
of it has involved steps that call for justification by their consequences, 
and that is the spirit in which we are proceeding here. In Ref. \cite{MP02}, 
which seems the most economical derivation possible, the point of view 
is taken that we smooth the fields (\eg, the velocity field ${\bf v}$ and, 
in MHD, the magnetic field ${\bf B}$) but not their ``sources'' ({\eg}, 
the vorticity field $\bom$ and, in MHD, the electric current 
density ${\bf j}$). By ``sources,'' we mean here the curls of 
${\bf v}$ and ${\bf B}$ which, given a set of boundary conditions, 
determine them through the Biot-Savart law or solutions to Poisson's 
equation. The assumption is that the large-scale dynamics are relatively 
insensitive to small changes in the positions of the sources, but are 
sensitive to the strengths and approximate positions of them. Said 
another way, the large-scale fields alone are assumed to be responsible 
for those motions of their sources which significantly affect those 
large-scale fields. This assumption is by no means self-evident, but does 
have the advantage of reducing the derivation to a single algebraic 
step, in contrast to some more involved derivations which have been 
given, and which seem logically no more compelling. Our focus here is 
on neither presenting unarguable alpha-model derivations or comparing 
the possible variants of it, but rather on exploring the consequences 
of the primitive version of it given here [Eqs. (\ref{E11}) and (\ref{E12}) 
in what follows].

Modeling MHD flows with the Lagrangian-averaged methodology has been barely
explored with no emphasis on the turbulence regime.
This article thus focuses on several such predictions, numerically obtained, 
for the case of two-dimensional magnetohydrodynamics, or hereafter 2D  MHD
(see \eg for a brief review, Ref. \cite{sanmin}). Several effects have 
been studied phenomenologically, theoretically and numerically
in the past, and may be identified in the 
literature by the names “selective decay,” “dynamic alignment,” direct and 
inverse “cascades,” and the characterization of probability distribution 
functions (pdf’s) for the fluctuating field variables. There are several 
Reynolds-like numbers which can be attributed to MHD turbulent flows, since 
there are two 
possible velocities which may appear in the numerators (the flow speed and the 
Alfv\'en speed) and two diffusivities that may appear in the denominators 
(kinematic viscosity and magnetic diffusivity). Some length scale 
characteristic of the initial fields is usually present in the numerators.  
All these Reynolds-like numbers can be made to appear in the places of 
reciprocals of the transport coefficients in front of the dissipative terms in 
various dimensionless representations of the MHD equations. In general, the 
larger the values of these Reynolds-like numbers (or equivalently, the smaller 
the transport coefficients), the greater the required numerical resolution to 
follow their solutions. Values of a Reynolds number like $10^4$ usually strain 
available computer 
resources even in two space dimensions (2D), and while the attainable total 
number of degrees of freedom with computers has been steadily 
increasing over several decades, 
there are situations in which one might be curious about results in cases of 
far higher values of direct interest for geophysical flows and yet not 
attainable in the foreseeable future. The alpha model, 
if it can be verified to give correct predictions in the range of accurate, 
un-modelized solutions, will acquire a certain credibility in providing the 
behavior (at least of the long-wavelength Fourier components) in situations 
with Reynolds-like numbers so high as to put them presently far out of reach 
of direct numerical solutions (DNS), particularly so in three space dimensions
(3D). Another set of regimes where modeling is needed is when widely disparate 
time and length scales occur, such as for either a small or large magnetic 
Prandtl number;
this is the case for the former in liquid metals as encountered in laboratory 
dynamo experiments and in breeder reactors, in the core of the earth and 
planets, and in the convective zones of the sun and stars, or for the latter 
in the interstellar medium.

\begin{figure}
\includegraphics[width=9cm]{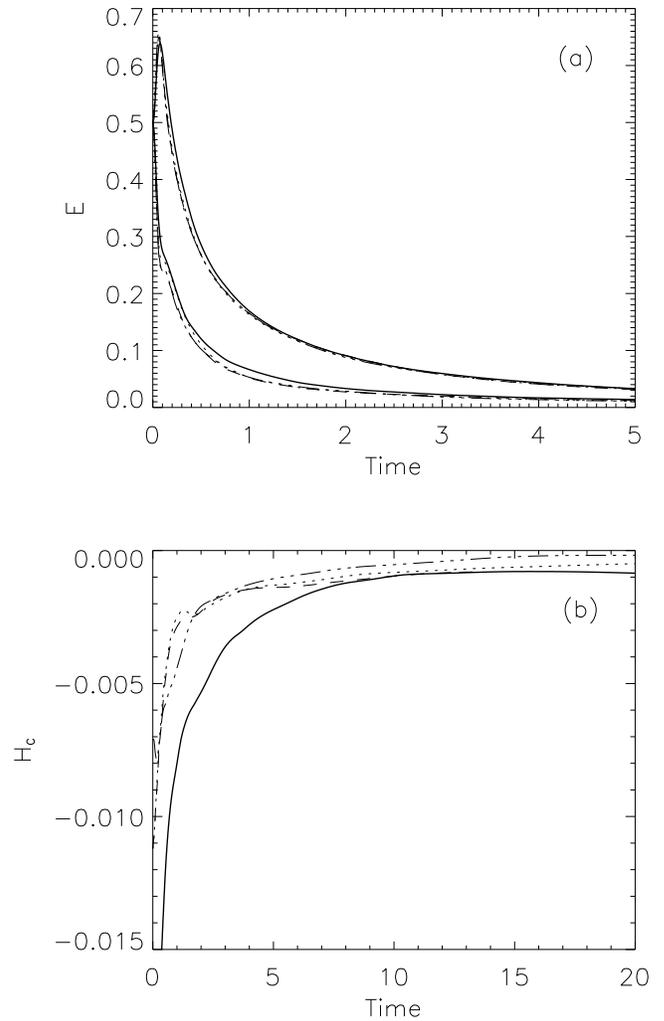}
\caption{(a) Magnetic energy (upper curves) and kinetic energy (lower curves) 
     as a function of time until $t=5$, and (b) cross helicity as a function 
     of time until $t=20$, for the selective decay runs 1-4 (see Table 
     \ref{tabr}). The temporal evolutions of DNS and alpha runs appear 
     similar.}
\label{fig1} \end{figure}
                                         
In Section II, we write down the alpha-model equations that we assume 
for incompressible, one-fluid  MHD with a minimum of theoretical 
justification, following Ref. \cite{MP02}; they include the effects of 
true viscous and dissipation. We provide expressions for ideal invariants 
that are conserved by the alpha model when the viscous and Ohmic 
dissipation coefficients are dropped,
and decay laws for them when the dissipation coefficients are present and 
finite. Much of the approach to and vocabulary of the way turbulence problems 
have historically 
been formalized for MHD (a subject where computational data vastly exceeds 
experimental or observational data) can by now be taken for granted. 
In Section III, we describe results for the problem of selective decay in 2D. 
In Section IV, we turn our attention to that of dynamic alignment of the 
velocity and magnetic fields in turbulent decays. In Section V, we focus on 
inverse cascade computations. 
In all these cases, there are comparisons to be 
made between the alpha-modeled results and direct solutions of the primitive 
MHD equations. In Section VI, we address ourselves to the problem of 
quantitatively assigning errors to the alpha model, as compared with 
well-resolved full MHD as well as with unresolved MHD, of resolution 
comparable to that used for the alpha model. In Section VII, we sample a few 
effects at Reynolds-like
numbers that are too high for any immediately foreseeable DNS code to 
approach. Finally, in Section VIII, we briefly summarize the results 
and suggest future problems in which the alpha model may have some utility.

To anticipate the conclusions of the paper, we note some features of the DNS 
solutions that the 
alpha model apparently finds out of reach. For instance, the location, in the 
plane, of specific features of evolving turbulent fields such as contour plots 
of vorticity or vector potential are virtually never accurately reproduced 
after short times (contours of constant magnetic vector potential in 2D
are magnetic field lines).
Likewise, the spectral details at the small scales are not accurate and are 
not expected to be, since it is modifications of the dynamics at small scales 
that make the alpha model possible in the first place. But as far as the long 
wavelength component behavior for the turbulent kinetic and magnetic spectra 
is concerned, the alpha model seems to recover the main features of MHD
turbulent flows in two space dimensions.

One technical feature of the computations peculiar to two-dimensional 
(2D) MHD should be commented upon. Because of the inherent tendency of 2D 
magnetic excitations to migrate to longer wavelengths, treatment of initial 
value problems requires beginning with excitations located in intermediate 
length scales, rather than at the longest wavelengths. In wavenumber space, 
this means that any filtering that is done must be done above wavenumbers 
corresponding to shorter wavelengths than those in the initial conditions. 
For this reason, very low resolution alpha-model calculations are not 
possible, unless the initial conditions themselves are to be left outside 
the basic box in Fourier space. This contrasts with the situation in 
three-dimensional hydrodynamics, where the emphasis is typically on 
cascades to shorter length scales, and where for the above reasons, it 
is possible to attempt large eddy simulations (LES) with very low 
maximum wavenumbers, and the initial excitations may all reside at 
the largest scales. While the ratio of our maximum retained wavenumber 
to the wavenumber where the filtering begins is about 8, it should be 
noted that larger ratios are feasible for three-dimensional Navier-Stokes 
LES and alpha model computations.

\section{THE ALPHA MODEL FOR MHD}
\subsection{The equations}

We write the three dimensional version of the equations first for the 
primitive incompressible MHD equations, and then for the alpha model.
We will then specialize them to two dimensions for the 
purposes of this paper. The basic variables are the velocity field $\bv$ 
and the magnetic field $\bB$, functions of space and time coordinates 
$(\bx,t)$. In dimensionless (“Alfv\'enic”) units, the equations are:

\begin{eqnarray}
&&\frac{\partial {\bf v}}{\partial t}+ {\bf v} \cdot \nabla {\bf v} =
-\nabla {\cal P} + \bj \times \bB -\nu \nabla \times \bom \label{E1} \\
&&\frac{\partial {\bf B}}{\partial t}+ {\bf v} \cdot \nabla {\bf B} = 
\bB \cdot \nabla {\bf v}
-\eta \nabla \times {\bf j} \label{E2}  
\end{eqnarray}
together with 
${\bf \nabla} \cdot {\bf v} =0$ and $\nabla \cdot {\bf B} =0$.

The velocity field may be considered to be expressed in units of an \rms value 
of the initial fluctuating velocity field, which we typically take to be 1. 
The magnetic field is made dimensionless by solving for the magnetic field 
value that would lead to an Alfv\'en speed equal to the \rms velocity field 
and dividing the magnetic field in laboratory units by that. The mechanical 
pressure is ${\cal P}$, which has first been divided by the mass density and 
then expressed in the units of the dimensionless velocity. The mass density is 
assumed to be constant and uniform. The viscosity $\nu$ and the magnetic 
diffusivity $\eta$ can be considered to be reciprocals of the mechanical and 
magnetic Reynolds numbers, respectively, in these units. Anticipating that the 
computations will be carried out inside a periodic box of edge $2\pi$,
the unit of length will in general be taken to be equal to unity, or about 
$1/6$ of a box dimension.

\begin{figure*}
\includegraphics[width=12cm]{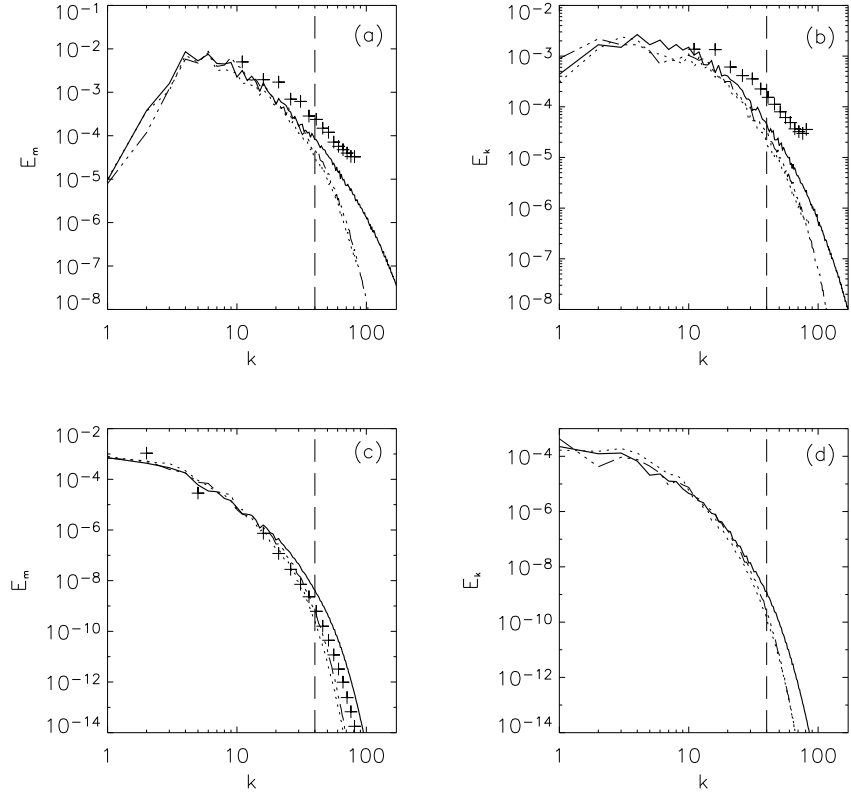}
\caption{(a) Magnetic energy and (b) kinetic energy spectra, for 
     selective decay runs 1-4 (see Table \ref{tabr}), at $t=5$; and (c) 
     magnetic energy and (d) kinetic energy spectra at $t=100$. The 
     vertical line gives  $k_{\alp}\sim 1/\alp$. The crosses ($+$) 
     indicate the place on the spectrum where an under-resolved DNS 
     run ($256^2$ grid points) departs significantly from the resolved 
     computed DNS spectra. This convention will also be used in 
     subsequent figures: crosses always indicate part of the 
     under-resolved DNS spectrum for the same initial conditions and 
     times.}
\label{fig2}
\end{figure*}
                                        
In the dimensionless units, the curl of the velocity is $\bom$, the 
vorticity field, and the curl of the magnetic field is $\bj$, the electric 
current density. The magnetic field $\bB$ can be written as the curl 
of a vector potential $\bA$, which, removing a curl from Eq. (\ref{E2}), 
obeys

\be
\frac{\partial {\bf A}}{\partial t}
= \bv \times \bB - \eta \bj - \nabla \Phi
\label{E3}\ee
where the scalar potential is $\Phi$. $\Phi$ can be determined by taking the 
divergence of Eq. (\ref{E3}), imposing the Coulomb gauge on $\bA$ (\ie
writing $\nabla \cdot \bA=0$), and solving the 
resulting Poisson equation for $\Phi$, involving $\bv$ and $\bB$ in the source 
term. (In a similar way, the pressure ${\cal P}$ can be found by taking the 
divergence of Eq. (\ref{E1}), using the vanishing of the divergence of the
time derivative of the velocity field $\bv$,
and solving the resulting Poisson equation for the pressure. These Poisson 
solutions are easy to solve in Fourier space.)

\begin{figure}
\includegraphics[width=9cm]{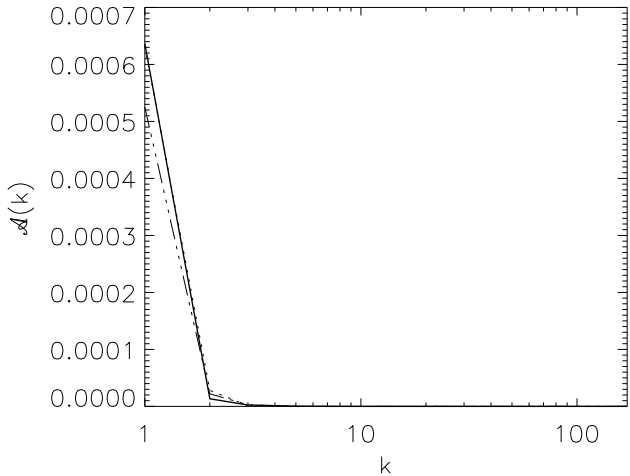}
\caption{Square vector potential energy spectrum shown at $t = 300$, for the 
      same selective decay runs as in Fig. \ref{fig2}.
      By this late time, essentially all of ${\cal A}$ is concentrated in 
      $k=1$.}
\label{fig3a} \end{figure}

To obtain the alpha-model version of Eqs. (\ref{E1}-\ref{E3}),
we may divide $\bv$ and 
$\bB$ into smoothed values plus fluctuations about those values. Thus

\be
\bv = \bus + \delta \bv
\label{E4}\ee
and
\be
\bB = \bBs + \delta \bB \ .
\label{E5}\ee

Here, the smoothed values of the fields, $\bus$ and $\bBs$, are defined by

\be
\bus = \int d^3\bxp \ \frac{\exp [-|\bx-\bxp|/\alp] }
                    {4\pi \alp^2 |\bx - \bxp|}\ \bv(\bxp,t)\ ,
\label{E6}\ee

\be
\bBs = \int d^3\bxp \ \frac{\exp [-|\bx-\bxp|/\alp] }
                    {4\pi \alp^2 |\bx - \bxp|}\ \bB(\bxp,t)\ ,
\label{E7}\ee
with $\alp$ at this point an arbitrary length; $\alp^{-1}$ is typically to be 
chosen as larger than the wavenumbers whose behavior it is desired to 
reproduce accurately.

In general, we use the same values of $\alp$ for smoothing $\bv$ and $\bB$, 
though we note the possibility of assigning {\it a priori} one value for 
$\bv$ and a different value for $\bB$ (respectively, $\alp_k$ and $\alp_m$).
Another possibility is to choose $\alp_m=0$ for 
$\bB$, which in that case will leave us with an unsmoothed magnetic field.
We will show such an example later.

We now take the curl of Eq. (\ref{E1}) to obtain the equation of motion 
in the vorticity representation,

\be
\frac{\partial \bom}{\partial t} + {\bf v} \cdot \nabla \bom=
 \bom \cdot \nabla {\bf v} + \nabla \times (\bj \times \bB) + \nu \nabla^2 \bom
\label{E8}\ee
and then substitute into Eqs. (\ref{E1}) and (\ref{E8}) the fields expressed 
in Eqs. (\ref{E4}) and (\ref{E5}). 
Note that we do not smooth the vorticity $\bom$ which can be regarded 
as the “source,” in a Poisson or Biot-Savart sense, of $\bv$. 
Nor do we smooth $\bj$, 
which bears the same mathematical relation to $\bB$ as $\bom$ does to $\bv$. 
The result is:
\begin{widetext} \be
\frac{\partial \bom}{\partial t} + (\bus+\delta \bv)\cdot \nabla \bom =
\bom\cdot \nabla (\bus+\delta \bv)+\nabla \times [\bj \times (\bBs+\delta \bB)]
+ \nu \nabla^2 \bom
\label{E9}\ee
and 
\be
\partial_t (\bBs+\delta\bB) + 
(\bus+\delta \bv)\cdot \nabla (\bBs+\delta\bB) =
(\bBs+\delta\bB) \cdot \nabla (\bus+\delta \bv) - \eta \nabla \times \bj \ ,
\label{E10}\ee \end{widetext}
upon which no approximations have as yet been made. That is, they are 
equivalent to Eqs. (\ref{E1})-(\ref{E2}).

Taking a modeling or heuristic point of view \cite{MP02}, the essence of the 
alpha model 
is to neglect the fluctuations $\delta \bv$ and $\delta \bB$ in relation 
to the smoothed fields $\bus$ and $\bBs$ in Eqs. (\ref{E9}) and 
(\ref{E10}), while leaving the source terms 
$\bom$ and $\bj$ alone. This is one way of looking at the alpha approximation. 
Its relation to other, more complicated derivations will not be discussed 
here, since our intent is to test the alpha model rather than to justify 
it from anything like first principles. Further discussion of the above 
approximation, which is the only one in our formulation, can be expected 
elsewhere.

The alpha model equations are then (removing a curl from Eq. (\ref{E9})):
\be
\partial_t \bv + \bus\cdot \nabla \bv + v_j \nabla u_s^j = -\nabla {\cal P}
+ \bj \times \bBs - \nu \nabla \times \bom
\label{E11}\ee
and
\be
\partial_t \bBs + \bus\cdot \nabla \bBs =
\bBs \cdot \nabla \bus- \eta \nabla \times \bj \ .
\label{E12}\ee
Note that the smoothed quantities bear the subscript letter $s$, and the 
unsmoothed ones do not. We shall follow this convention throughout. 
Eq. (\ref{E12}) could be viewed alternatively as a hyper-resistivity 
approximation on $\bBs$. The connection between the smoothed and unsmoothed 
fields may be stated in differential form as
\be
\bv = (1-\alp^2 \nabla^2)\ \bus
\label{E13}\ee
and
\be
\bB = (1-\alp^2 \nabla^2)\ \bBs \ .
\label{E14}\ee
We may associate smoothed values of $\bom$ and $\bj$ with the unsmoothed ones 
according to the same recipe; even though they do not enter directly into the 
dynamical equations, they are at some points convenient to think and talk 
about.
Thus $\nabla \times \bus \equiv \boms$, similarly 
$\nabla \times \bBs \equiv \bjs$,
and $\nabla \times \bAs \equiv \bBs$. A smoothed vector potential $\bAs$
may be regarded as having a curl $\bBs$, 
while obeying a Poisson relation to $\bjs$, namely $\nabla^2 \bAs = -\bjs$.
We stress that $\boms$ and $\bjs$ do not enter the alpha model equations 
we use.

Specialization to two dimensions is achieved by taking the curl of 
Eq. (\ref{E1}) or specializing Eq. (\ref{E1}) and Eq. (\ref{E3})
to the two dimensional geometry in
which there are only two $(x,y)$ non-zero components of $\bv$ and $\bB$, 
and only one component ($z$) of $\bom$ or $\bj$, and carrying out the 
smoothing approximations so described.
All fields are independent of the $z$ coordinate. Noting that only one 
component of $\bA$, the z-component, is relevant to two dimensions, the 
result is:

\be
\frac{\partial \omega}{\partial t} + \bus \cdot \nabla \omega = 
\bBs \cdot \nabla j + \nu \nabla^2 \omega \ ,
\label{E15}\ee

\be
\frac{\partial A_{s_z}}{\partial t} + \bus \cdot \nabla A_{s_z} = 
- \eta j \ ,
\label{E16}\ee
where there are stream functions $\psi$ and vector potentials $A_z$ that bear 
Poisson relations to their sources, both for the smoothed and 
unsmoothed versions:

\be
\nabla^2 \Psi = - \omega \ , \ \ \ \ \  \nabla^2 \Psi_s = - \omega_s \ ,
\label{E17}\ee

\be
\nabla^2 A_z= - j \ , \ \ \ \ \  \nabla^2 \Asz = - j_s \ .
\label{E18}\ee
To re-iterate, the principal intent of this paper is to compare typical 
solutions of Eqs. (\ref{E15}) and (\ref{E16}) with solutions, for the same 
initial and boundary conditions, of the well-known 2D MHD equations, 
unsmoothed.

\begin{table}
$$\vbox{\offinterlineskip\halign{\tv#&\cc{#}&
\tv#&\cc{#}&\tv#&\cc{#}&\tv#&\cc{#}&\tv#&\cc{#}&\tv#&\cc{#}&\tv#\cr
\noalign{\hrule}
\traithorizontal \traithorizontal
\tvi&Run&&$\alpmm$&&$\alpmv$&& N&&$\Rla$&& Figs. &\cr
\traithorizontal \traithorizontal
\tvi& 1&&$\infty$&&$\infty$&&1024&&215&& \ref{fig1}--\ref{fig4} &\cr
\traithorizontal
\tvi& 2&&40&&40&&1024&&235&& $- - -$ &\cr
\traithorizontal
\tvi& 3&&40&&40&&512&&240&& $-\cdots-$ &\cr
\traithorizontal
\tvi& 4&&40&&40&&256&&240&& $\cdots\cdots$ &\cr
\traithorizontal \traithorizontal
\tvi& 5&&$\infty$&&$\infty$&&512&&280&&  \ref{fig5}--\ref{fig8} &\cr
\traithorizontal
\tvi& 6&&20&&20&&512&&305&& $- - -$ &\cr
\traithorizontal
\tvi& 7&&20&&20&&128&&300&& $\cdots\cdots$ &\cr
\traithorizontal \traithorizontal
\tvi& 8&&$\infty$&&$\infty$&&256&&28 &&  \ref{fig9}--\ref{fig13} &\cr
\traithorizontal
\tvi& 9&&20&&20&&256&&30&& $- - -$ &\cr
\traithorizontal
\tvi&10&&30&&30&&256&&29&& $-\cdot-$ &\cr
\traithorizontal
\tvi&11&&$\infty$&&20&&256&&28&& $-\cdots-$ &\cr
\traithorizontal
\tvi&12&&30&&30&&128&&30&& $\cdots\cdots$ &\cr
\traithorizontal \traithorizontal
\tvi&13&&$\infty$&&$\infty$&&1024&&1,150&& \ref{figdiss}--\ref{figE3} &\cr
\traithorizontal
\tvi&14&&50&&50&&256&&1,170&& $- - -$ &\cr
\traithorizontal
\tvi&15&&$\infty$&&50&&256&&940&& $-\cdots-$ &\cr
\traithorizontal
\tvi&16&&$\infty$&&$\infty$&&256&&980&& $\cdots\cdots$ &\cr
\traithorizontal \traithorizontal
\tvi&17&&300&&300&&2048&&5,200&& \ref{figreyt}--\ref{figreys} &\cr
\traithorizontal \traithorizontal
\noalign{\hrule}}}$$
\caption{Main characteristics of the runs.
   $\alp_k^{-1}$ and $\alp_m^{-1}$ are the
   reciprocal of the alpha lengths for the velocity and the magnetic field;
   $N$ is the grid resolution before dealiasing, $\Rla$ is the Taylor
   Reynolds number (see Eq. \ref{taylor_r}) at peak dissipation,
   and the last column gives the 
   figures relating to the different runs, namely:
   runs 1--4 for selective decay, runs 5--7 for dynamic alignment,
   runs 8--12 for the inverse cascade of magnetic potential and
   runs 13--17 for large-scale turbulence. In
   the figures, solid lines are for fully
   resolved DNS (runs 1,\ 5,\ 8 \& 13),
   dashed lines for runs 2,\ 6,\ 9 \& 14, dashed-triple dots for runs 
   3, 11 \& 15, dotted lines for runs 4,\ 7,\ 12 \& 16, and a dash-dot 
   line for run 10.}
\label{tabr} \end{table}

\subsection{The invariants}
There are three ideal invariants for both sets of equations: the total 
energy $E$, the total cross helicity $H_C$, and the total mean-square vector 
potential ${\cal A}$.
The alpha model expressions for these are, respectively, 

\be
E = \frac{1}{2} \int d^2\bx\ (\bus \cdot \bv + \bBs \cdot \bB) \ ,
\label{E19}\ee

\be
H_C = \frac{1}{2} \int d^2\bx\ \bv \cdot \bBs \ ,
\label{E20}\ee
and

\be
{\cal A} = \frac{1}{2} \int d^2\bx\ \Asz^2\ .
\label{E21}\ee
Note that the energy invariant $E$
involves both the smoothed and unsmoothed velocity and magnetic field,
whereas only the smoothed magnetic variables appear in the expressions of
$H_C$ and ${\cal A}$, 
due to the linearity of the induction equation, once the velocity field is 
given; however, the decay rates of $H_C$ and ${\cal A}$ involve both the 
smoothed and unsmoothed magnetic variables whereas the decay rate of energy 
only involves the unsmoothed current density (see below).

The decay laws for these quantities are, in periodic boundary conditions, 
\be
\frac{dE}{dt} = - \nu \int d^2\bx \ \omega \omega_s 
- \eta \int d^2\bx \ j^2 \ ,
\label{E22}\ee
\be
\frac{dH_C}{dt} = 
-\frac{1}{2} \eta \int d^2\bx \ \omega j
-\frac{1}{2} \nu \int d^2\bx \ \omega j_s  \ ,
\label{E23}\ee
and
\be
\frac{d {\cal A}}{dt} 
= - \eta  \int d^2\bx \ \Asz j 
= - \eta  \int d^2\bx \ \nabla \Asz \cdot \nabla A_z \ .
\label{E24}\ee

These three invariants will be at the core of the numerical tests 
to be reported in the next three Sections. Finally, it
should be noted that Eqs. (\ref{E19})-(\ref{E24}) differ from their full
MHD equivalents in detail, but approach them as $\alp \rightarrow 0$.

For convenience and later reference, the main characteristics of all the runs 
described in this 
paper are given in Table \ref{tabr}, together with the number of the
figures related to the different category of runs. $N$ is the grid resolution
before dealiasing, using the standard $2/3$ rule
in all the runs described in this paper
(hence, for a run on a grid of $N\times N$ points, 
the maximum wavenumber attainable is equal to $N/3$). Finally,
$\Rla$ is the Taylor Reynolds number defined as
\be
R_{\lambda}=\frac{\lambda v_{rms}}{\nu} \ ;
\label{taylor_r}\ee
it is based on the \rms velocity and on the Taylor scale 
\be
\lambda=2\pi\sqrt{<v^2>/<\omega^2>} \ ,
\label{taylor_s}\ee
computed at the peak of the dissipation.
The viscosity is equal to $5\times 10^{-4}$ for the selective decay runs
1--4, with initial conditions with non-vanishing Fourier coefficients 
(see Eq. (\ref{E25})) for
wavenumbers between $k_1=10$ and $k_2=30$.
Runs 5--7 are for dynamic alignment, with $\nu=10^{-3}$ and
$k_1=5,\ k_2=10$.
Runs 8--12 are for the inverse cascade of magnetic potential, with
$\nu=10^{-3}$, the forcing occurring in the interval $k_1=18,\ k_2=22$.
Runs 13--16 deal with large-scale turbulence with
$\nu=5\times 10^{-4}$ and the initial conditions confined between
$k_1=1$ and $k_2=3$, whereas for run 17, $\nu=2\times 10^{-5}$.  
In the figures, solid lines are for fully
resolved DNS (runs 1,\ 5,\ 8 \& 13),
dashed lines for runs 2,\ 6,\ 9 \& 14, dashed-triple dots for runs with
$\alp_k\not= 0,\ \alp_m\equiv 0$ (runs 3, 11 \& 15); finally, dotted lines
are for runs 4,\ 7,\ 12 \& 16, and a dash-dot line for run 10.
Note that all runs have unit magnetic Prandtl number; the initial 
conditions are such that kinetic and magnetic energies are equal and of order
unity with random phases. All runs are decaying (\ie no forcing), except for 
the inverse cascade runs, which have 
zero initial conditions and forcing in the induction equation only.

\begin{figure*}
\includegraphics[width=12cm]{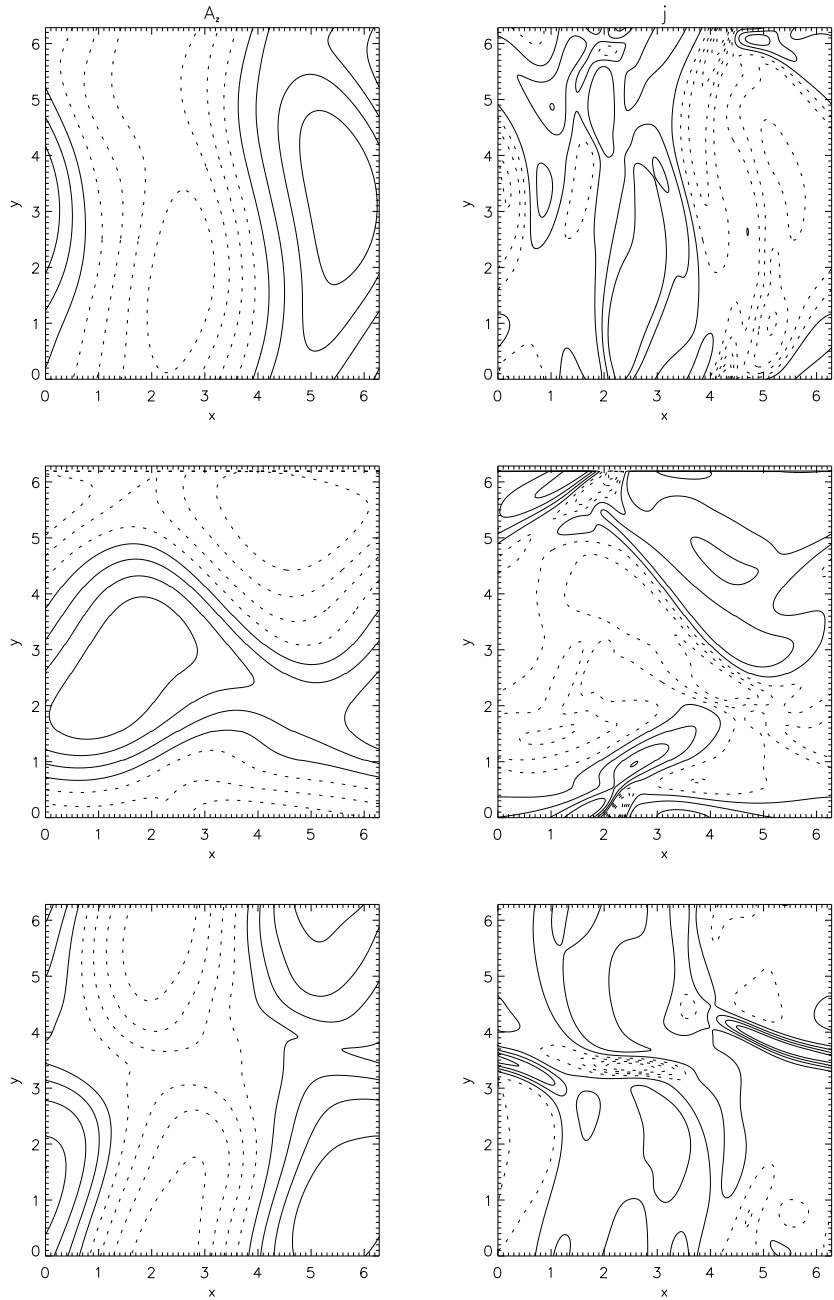} 
\caption{Contour plots of vector potential (left) and current density (right) 
    at $t=400$ for the DNS run (top) and alpha run 3 (middle) and 4 (bottom),
    with positive and negative values respectively in solid and dashed lines.
    Large scales are almost in the form of bars parallel to the axes.}
\label{fig3} \end{figure*}

\begin{figure}
\includegraphics[width=9cm]{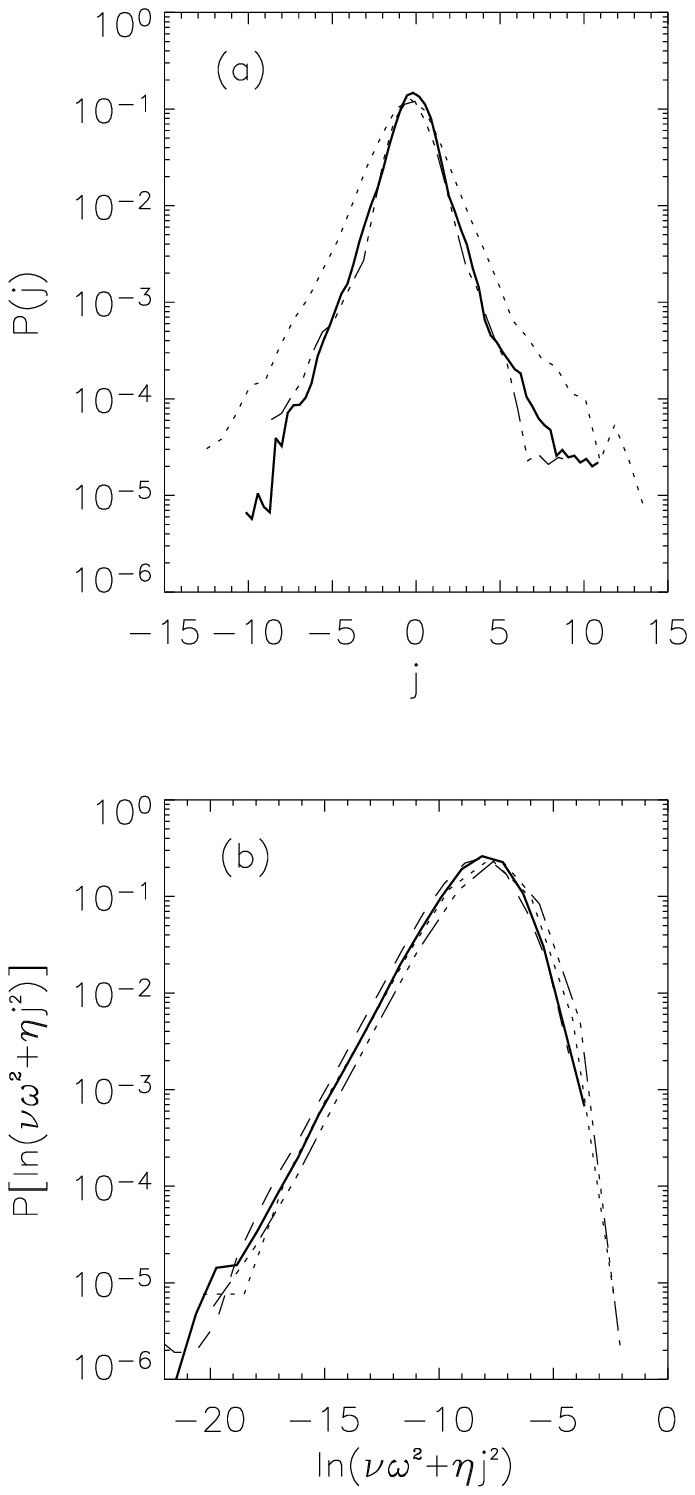}
\caption{Pdfs at $t = 15$ of (a) the current density, and (b)
    $\ln(\nu \omega^2+\eta j^2)$,
    for all selective decay runs (see Table \ref{tabr}).
    Note the larger values of current density for the alpha runs compared to 
    the DNS (solid line).}
\label{fig4} \end{figure}

\section{SELECTIVE DECAY}

By ``selective decay,'' we mean turbulent processes 
(see \eg \cite{MM80}-\cite{kmwt95}) in which one or
more ideal invariants are dissipated rapidly relative to another, due to the
transfer of the dissipated quantities to short wavelengths where the
dissipation coefficients become effective. In 2D MHD, with negligible
cross-helicity, the selectively dissipated quantity is energy, while the
nearly-conserved quantity is mean square vector potential. The limit defines a
variational problem which seeks the state in which the dissipated quantities
are as close to zero as they can be for the surviving value of the
nearly-conserved quantity. There are no constraints on the cascade of kinetic
energy to short wavelengths, so the asymptotic state is expected to be one
that is largely magnetic and has the surviving magnetic excitations peaked at
the longest wavelengths. In particular, the vector potential spectrum should
have a sharp maximum at the lowest wavenumber of the computation, here
$k_{min}=1$. This effect has been demonstrated repeatedly in the past
(\cite{MM80}-\cite{kmwt95}).

In this Section, we compare the full MHD ($\alpha \equiv 0$) results for a
selective decay run of a familiar type with the consequences of the alpha
model for the same initial conditions but finite $\alpha$. We specify the
initial conditions in Fourier space, with $\bv$ and $\bB$ represented as the
Fourier series,
\be
\bv(\bx,t)=\sum_{\bk} \hat \bv(\bk,t)e^{i \bk \cdot \bx} \ , \ 
\bB(\bx,t)=\sum_{\bk} \hat \bB(\bk,t)e^{i \bk \cdot \bx} \ ,
\label{E25}\ee
with similar decompositions for the other vector fields.
The non-vanishing initial Fourier coefficients are confined to a ring in
$\bk$-space between $k_1=10$ and $k_2=30$. The amplitudes in this ring are
chosen equal, for both $\bv$ and $\bB$, and the phases are chosen from a
random-number generator. The overall normalization is such that the initial
kinetic and magnetic energies are both $0.5$, referred to a unit 
(two-dimensional)
volume. The maximum value of $k$ is, after the de-aliasing, $k_{max}=341$ and
the time step is $\Delta t=5 \times 10^{-4}$; the dimensionless viscosity 
and resistivity are both equal to $5 \times 10^{-4}$;
the initial Reynolds numbers, kinetic and magnetic, are then 
formally equal respectively to $2000$, based on a unit
length scale in the basic box of edge 2$\pi$, whereas the Taylor Reynolds 
number at peak of dissipation is equal to $215$ for the DNS run, and slightly 
larger for the alpha runs (see Table \ref{tabr}).
The time will be measured in units that are defined by the ratio of
unit length to the initial \rms velocity; based on the energy containing scale,
one time unit can be several initial eddy turnover times, a number which may
increase or decrease as the kinetic energy is dissipated.

\begin{figure}
\includegraphics[width=9cm]{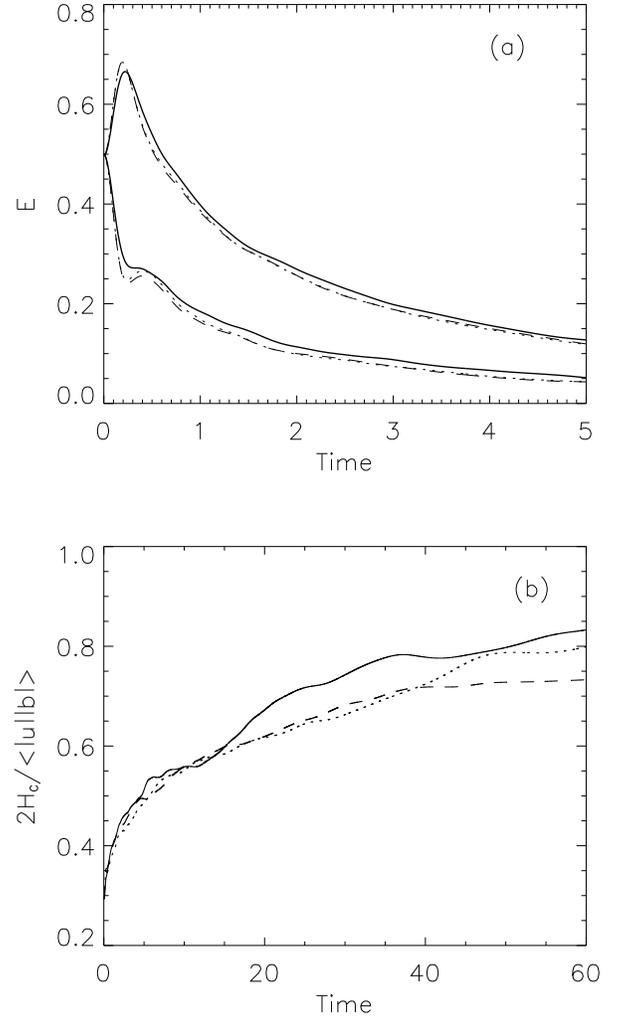}
\caption{(a) Temporal evolution of magnetic energy (top curves) and kinetic 
    energy (bottom curves), and (b) of normalized cross helicity, for all 
    dynamic alignment runs (see Table \ref{tabr}).
    Note that only the first five units of time for the run are shown in (a);
    the solid line is for the DNS run.}
\label{fig5} \end{figure}

\begin{figure*}
\includegraphics[width=12cm]{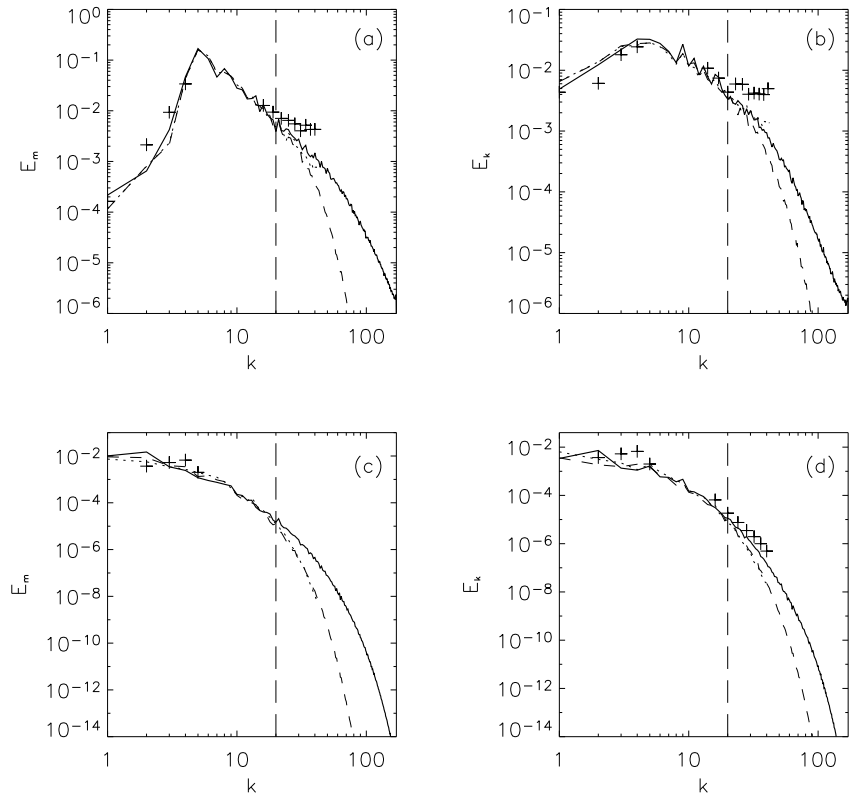}
\caption{(a) Magnetic and (b) kinetic energy spectra, for the three 
    dynamic alignment runs (see Table \ref{tabr}), at $t=2$; (c) 
    Magnetic and (d) kinetic energy spectra at $t=30$. As elsewhere, 
    crosses ($+$) indicate points on the computed under-resolved 
    DNS spectrum ($128^2$ grid points) for the same initial conditions, 
    and indicate the values of $k$ where the first significant departures 
    occur from the well-resolved DNS.}
\label{fig6} \end{figure*}

Figure \ref{fig1}a displays the computed magnetic energy (upper curves) and
kinetic energy (lower curves) versus time, showing the ultimate dominance of
the magnetic energy over the kinetic, by an order of magnitude at this time.

The solid lines are the results of the
full MHD computation (\ie, $\alpha =0$), with $1024^2$ grid points. The
dashed lines are the results of the alpha model with $1024^2$ grid points and
with $\alpha = 1/40$. The dotted line (barely distinguishable from the
dashed one) shows the results for an alpha-model run with $256^2$ grid points
and the same value of $\alpha$. Note that a full MHD run with the lower number 
of grid points, \ie an under-resolved computation of MHD turbulence, would 
display disagreement with the other three runs.
We shall discuss this question in Section VI.

Fig. \ref{fig1}b shows the cross helicity (which remains very small, relative 
to the energy, because the random phases for the two fields imply negligible
correlation between them) as a function of time. The alpha model computations
disagree with the “exact” MHD run, but since all the quantities are so small,
this disagreement is not deemed to be significant, but rather occurs as
fluctuations around values close to zero. We should remark that in
both cases, the comparison between the full MHD quantities and their
alpha-model analogues has been done after a rescaling of initial data
which makes the energies agree exactly at $t=0$ (with the full, unsmoothed 
MHD values); the original MHD energy is not quite the same
as the energy integrals defined in Section II involving the smoothed fields,
because of the smoothing, though the difference is only a few percent.

\begin{figure*}
\includegraphics[width=12cm]{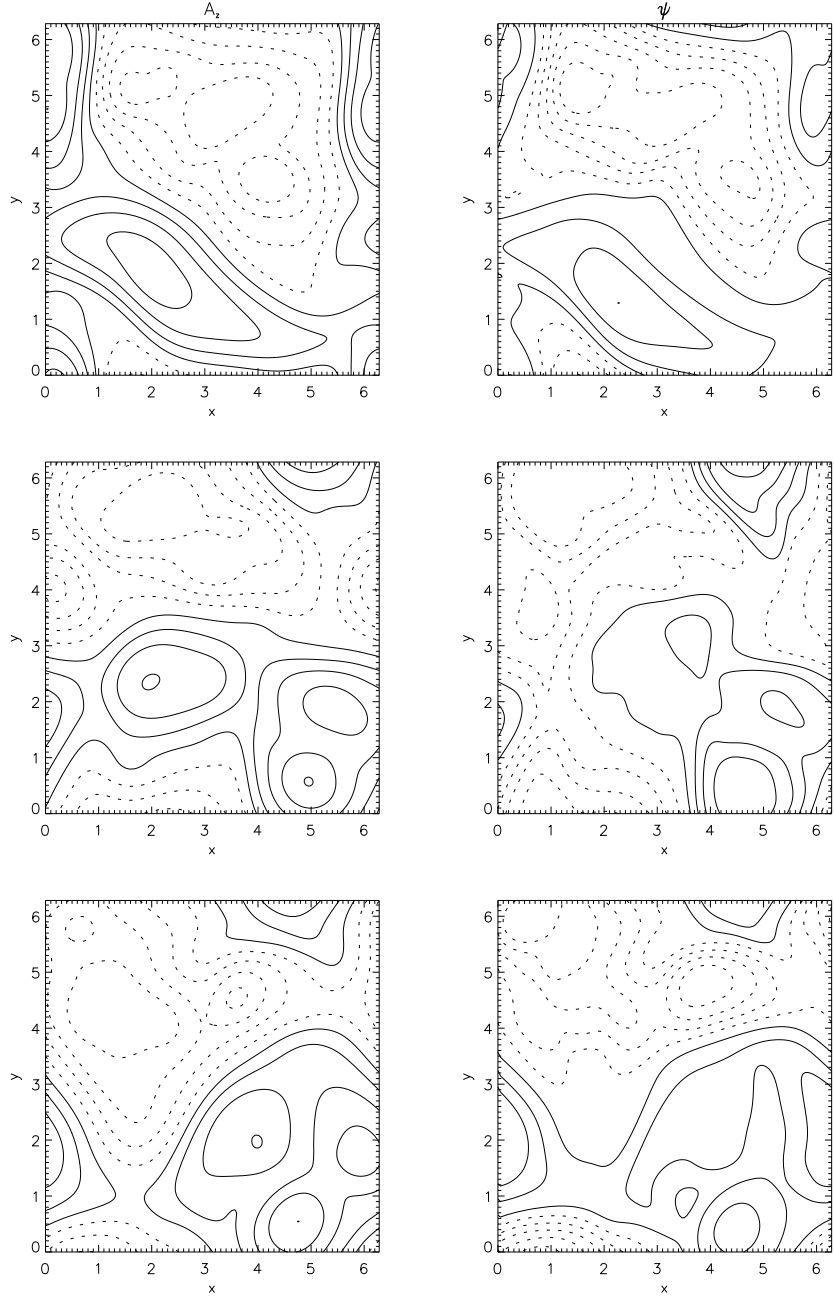} 
\caption{Right column: Contour plots of the (unsmoothed) stream function 
    at $t=60$ for dynamic alignment, with the DNS run at the top, run
    6 in the middle and run 7 at the bottom
    (see Table \ref{tabr}). Left column: contours of the 
    (smoothed) vector  potential at the same time.
    In all three runs, contours of magnetic potential and 
    stream function are similar by that time, more so for the DNS run,
    and concentrated in the large scales.}
\label{fig7} \end{figure*}

\begin{figure}
\includegraphics[width=9cm]{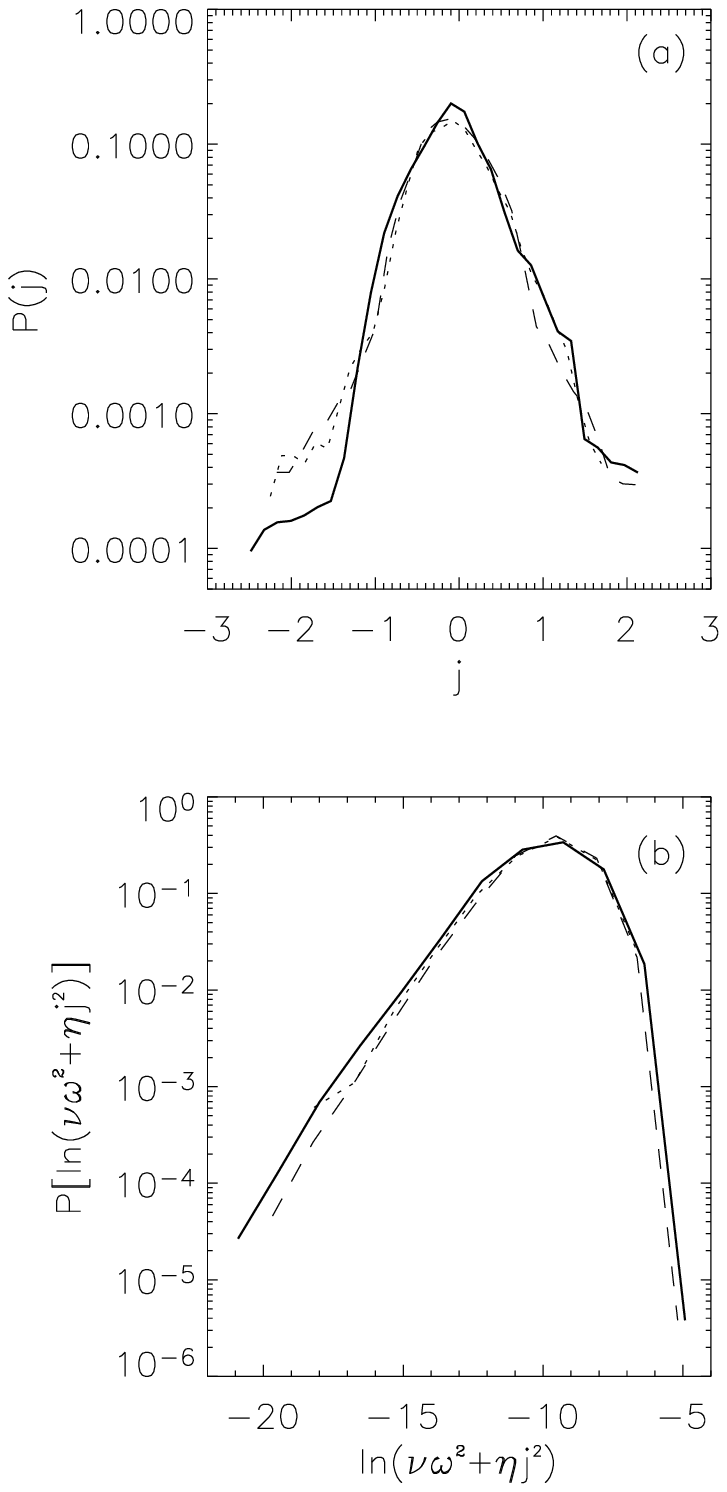}
\caption{Pdfs at $t = 50$ of (a) the current density, and (b) 
    $\ln(\nu \omega^2 + \eta j^2)$, for all dynamic alignment runs (see Table 
    \ref{tabr}). Note again the high values of $|\bj|$ for the alpha runs.}
\label{fig8} \end{figure}

\begin{figure}[h!]
\includegraphics[width=9cm]{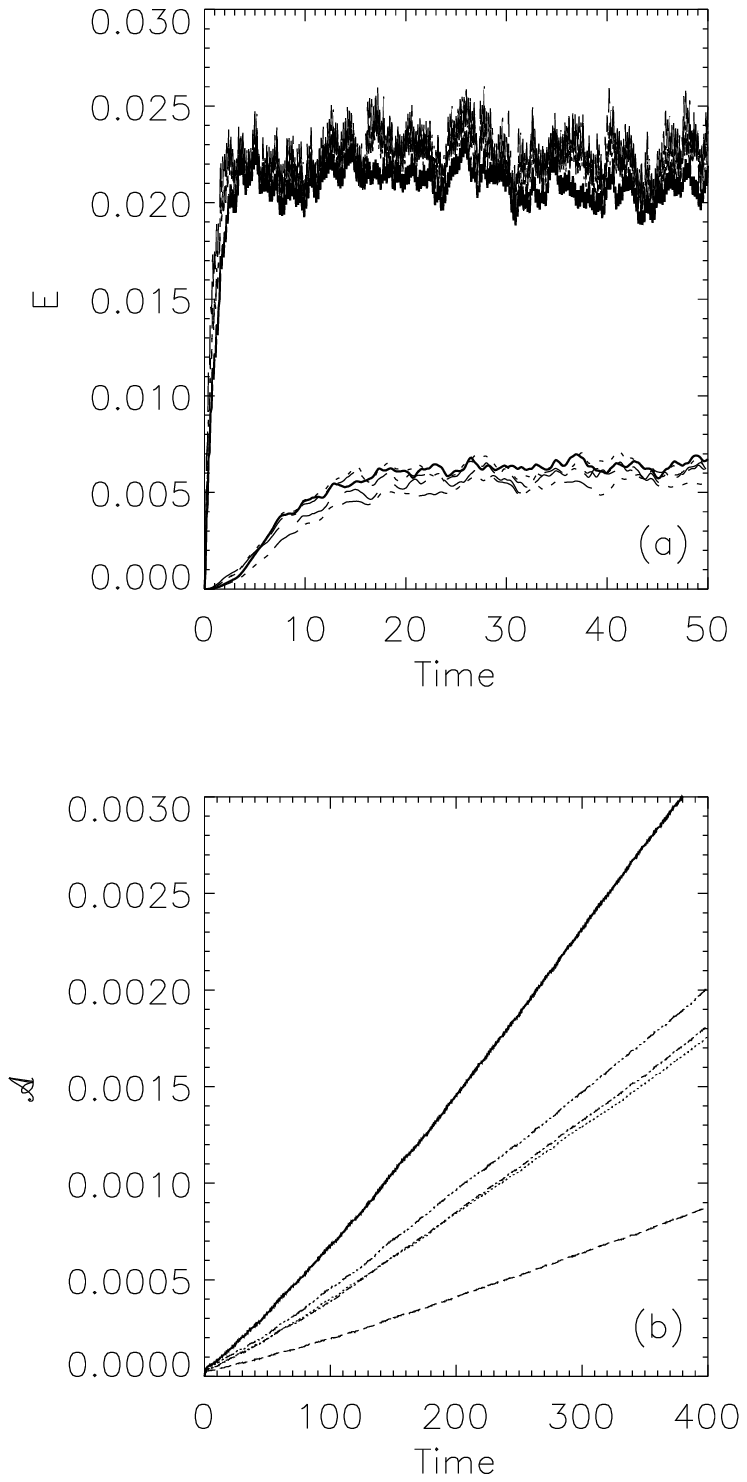}
\caption{(a) Temporal evolution of the magnetic energy (top curves) and 
    kinetic energy (bottom curves) until $t=50$, and (b) of the squared 
    vector potential until $t=400$, for all 
    inverse cascade runs (see Table \ref{tabr}, runs 8-12).
    Whereas energies are in good agreement with the DNS (shown as usual 
    with a solid line), the growth of squared 
    magnetic potential is slower for all alpha runs (see also 
    Fig. \ref{fig13}).}
\label{fig9} \end{figure}

Figures \ref{fig2}a,b show the omni-directional energy spectra for the 
magnetic and kinetic energies (as defined in Section II when $\alpha$ is 
non-zero), respectively, at $t=100$. The same conventions 
adopted in Fig. \ref{fig1} ({\it i.e.}, solid lines mean full MHD, 
dashed lines mean alpha model with
$\alpha = 1/40$ and $1024^2$ grid points, and dotted lines mean runs
done with the same value of $\alpha$ but with $256^2$ grid points) will be
followed throughout. The alpha-model spectra that are plotted result from
taking the spectral density of the invariants in
Equations (\ref{E19})-(\ref{E21}). The vertical line indicates the wavenumber
$k_{\alp}$ corresponding to the length $\alp$. Note that the 
under-resolved spectra begin to differ at $k\sim k_{\alp}/2$, but 
the $\alpha$-model and well-resolved DNS agree up to $k\sim k_{\alp}$.

Figure \ref{fig3a} displays a spectrum in log-lin scales, of the vector 
potential at very late times, when the selective decay is nearly complete and 
the magnetic excitations
are concentrated in the longest wavelength allowed by the boundary conditions
($k_{min}=1$). The alpha model has reproduced this feature, with only a small
disagreement in the values at $k=1$
(see also Section VI for a more complete discussion of errors).

The suppression of the small scales is quite apparent for
both quantities, but the large scales, like the global energies exhibited in
Fig. \ref{fig1}a, do not appear to be significantly affected. 
The lower resolution alpha model run reproduces the  same result.
This is what can be realistically hoped for from the alpha model, although we
note that the disagreement between the true MHD run and the alpha runs starts
at a scale roughly twice as large as $\alp^{-1}$.

Figures \ref{fig3} display contour plots of curves of 
constant vector potential $A_{s_z}$ (Fig. \ref{fig3}, left) 
and constant current density $j$ (Fig. \ref{fig3}, right) 
at time $t = 100$. The top panel is for full MHD, the middle one for 
$\alpha = 1/40$ and $512^2$ grid points, and the bottom one
for $\alpha = 1/40$ and $256^2$ grid points. The flow may evolve toward
a state reminiscent of those found in \cite{bar} in the
case of 2D Navier-Stokes turbulence, with structures parallel to 
either axis. While there are marked similarities in the kinds of structures 
present in the DNS and in the alpha runs, there are
clearly no one-to-one correspondences as to specific features, either as to
location, orientation, or intensity. From these and many similar figures we
have looked at, we have concluded that while the alpha model does an excellent
job of reproducing long-wavelength spectra, the pointwise details of the
solution are not well tracked by it, at least in the absence of constraining
material boundaries.

In Figs. \ref{fig4}, we display normalized probability distribution functions
(pdfs) of the current density $j$ in Fig. \ref{fig4}a, 
and in Fig. \ref{fig4}b of the spatial density of the dissipation rate of
energy given in Eq. (\ref{E22}); note that the local (spatial) dissipation of 
kinetic energy differs in its expression, involving the symmetrized velocity 
gradient instead of the local squared vorticity density.  The conventions
with the lines are the same as those in the preceding three figures. It is
apparent that the pdfs of the alpha model do a good job for the lower values
of $|\bj|$ but do not reproduce the tails accurately, in
particular at lower resolution, \ie
intermittency is not fully reproduced, and is not expected to be
(for a study of intermittency in the context of LES, see \eg Ref. \cite{LK01}).
The same is true of the dissipation density, although discrepancies appear 
smaller.

\section{DYNAMIC ALIGNMENT}

A perfectly ``aligned'' solution to the ideal version of Eqs. (\ref{E1}) and 
(\ref{E2}) results whenever $\bv = +\bB$ or $\bv = -\bB$ everywhere. Previous 
computations \cite{GPL83}-\cite{XX1}, inspired by observations in the quiet 
solar wind \cite{XX2}, 
have shown that MHD turbulence in which a significant degree of initial 
alignment, or correlation between the $\bv$ and $\bB$ fields, exists will 
evolve toward a state of greater and greater alignment as time goes on. The 
physical origins of this process are not completely clear, except that we 
may note that an aligned state involves no spectral transfer to higher wave 
numbers where viscous and Ohmic dissipation are effective, so that those 
patches where alignment exists initially may have a tendency simply to 
outlive the more active, unaligned patches where spectral transfer makes 
dissipation more likely. Similar alignment, this time between velocity 
and vorticity, can be observed for three-dimensional Navier-Stokes flows 
(see \eg Ref. \cite{tsinober} for an experimental study).

\begin{figure}
\includegraphics[width=8cm]{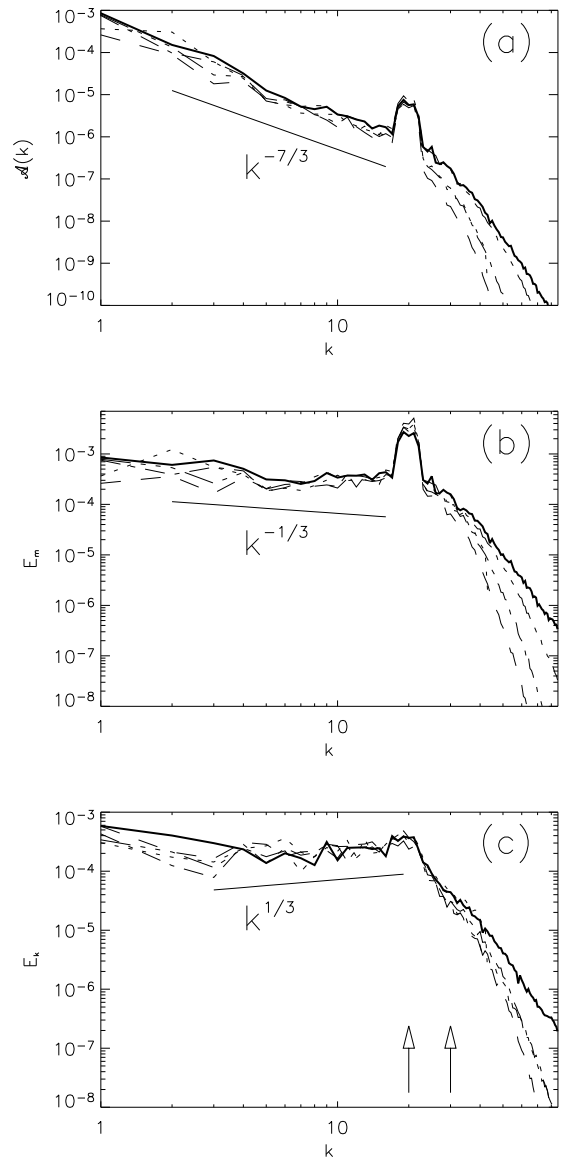} 
\caption{(a) Squared vector potential, (b) magnetic, and (c) kinetic 
    energy spectra, for all inverse cascade runs (see Table \ref{tabr}, 
    runs 8--12) at $t=160$, with as usual a solid line for DNS runs; 
    the two vertical arrows correspond to the two values of alpha. 
    The line with a $-7/3$ slope follows the phenomenological 
    prediction for the magnetic potential spectrum derived in 
    \cite{P78}. As $\alpha$ gets closer to the forcing wavenumbers,
    the inverse cascade is slowed down more significantly.}
\label{fig10} \end{figure}

A useful index of the global degree of alignment of $\bv$ and $\bB$ may be 
taken as the ``fractional alignment'': it is defined taking $2H_C$ and 
dividing it by the square root of $<\bus\cdot \bv><\bBs\cdot \bB>$, where 
angle brackets mean spatial averages. When this ratio is unity, the fields 
may be regarded as perfectly aligned. In Fig. \ref{fig5}, we display the 
results of a run which starts 
with the Fourier amplitudes chosen to be equal in a ring with $5\le k \le 10$,
with unit \rms values of $\bv$ and $\bB$ and with phases chosen so that the 
fractional alignment is initially 0.3. Finally, $\Delta t = 10^{-3}$ and 
$\nu = \eta = 10^{-3}$; the Reynolds numbers are equal to $1000$, based upon 
unit length, unit \rms velocity at $t=0$ and the transport coefficient, and the
Taylor Reynolds number at peak dissipation is equal to $280$ for the DNS run,
and again slightly higher for the alpha runs (see Table \ref{tabr}).

In Fig. \ref{fig5}a, we show that the energies, magnetic (top curves) 
and kinetic (lower curves), as functions of time, have comparable evolutions.
The fields, however, become progressively more aligned 
during their decay as can be seen in Fig. \ref{fig5}b, where the 
alignment index gradually increases from $0.3$ to about $0.85$. As 
before, the $\alpha = 0$, or full MHD (with $512^2$ grid points),
results are exhibited as solid lines, the dashed lines are the 
results for $\alpha = 1/20$ and $512^2$ grid points, while the dotted line 
is for the same alpha but with only $128^2$ grid points. The rather large 
discrepancies for long times for the correlation coefficient may come from the 
fact that it involves a normalization; the cross-helicity $H_C$ themselves do 
not differ significantly (not shown); it also may imply that small scales, 
which are modified by the alpha modeling process, play a role in the growth of 
large-scale correlations between the velocity and the magnetic field.

\begin{figure}
\includegraphics[width=9cm]{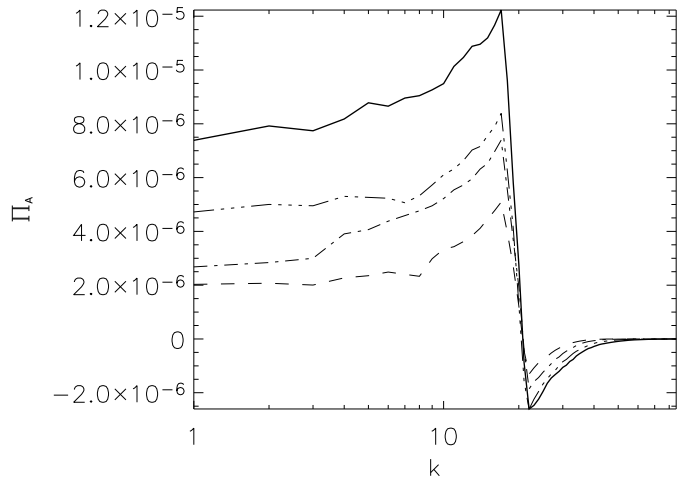}
\caption{Flux of squared vector potential in Fourier space $\Pi_A(k)$ 
     defined in Eq. (\ref{flux})
     at $t = 400$; same runs and symbols as in Fig. \ref{fig10}.
     Note the lesser amount of flux in the alpha runs compared to the 
     DNS (solid line).}
\label{fig11} \end{figure}

The spectra in Figs. \ref{fig6} show that in the aligning situation, the 
alpha model continues to do a good job of reproducing the long wavelength 
components of the MHD spectrum.
Figure \ref{fig7} shows contour plots of the unsmoothed stream function (right)
at $t=60$, for full MHD (top), for the alpha model with $\alpha = 1/20$ 
and $512^2$ grid points (middle), and for the alpha model with 
$\alpha = 1/20$ and $128^2$ grid points (bottom); Figs. \ref{fig7} (left)
show the corresponding contours for the smoothed vector potential.
The pointwise alignment is more visible in the top contours (\ie for the full
MHD run) than in the finite $\alp$ runs.

Figure \ref{fig8} shows pdfs of (a) the current density $\bj$, and (b) 
$\ln(\eta j^2 + \nu \omega^2)$ \ie the spatial density of the energy 
dissipation rate; the units scales are lin-log; note the exponential 
dependency at low values, as already found in \cite{biskamp2d}, and also 
visible in Fig. \ref{fig4}. 
As in Section III, the only significant disagreement occurs in the 
tails, and  the statistical properties of the turbulence are once again rather 
well reproduced by the alpha model approximation, even if the pointwise 
features shown in Fig. \ref{fig7} are not particularly well reproduced.

\section{INVERSE CASCADE OF VECTOR POTENTIAL}

One might guess that one of the more demanding tests of alpha-modeled 2D MHD 
would be its performance in an inverse cascade situation. The essence of the 
alpha model is that it suppresses the small scales to some degree. In direct 
cascade situations, the transfer is largely from the large scales to the 
small, and intuitively, it has often been reasoned that as long as the 
small scales can be made simply to disappear at that point, there should 
be no harm done in the large scales. Accurate or not, this reasoning is 
behind the use of Large Eddy Simulations (see \eg Ref. \cite{MK00} for a 
recent review), “eddy viscosities” and “eddy resistivities” that are 
sometimes employed  to decrease the amount of resolution required for 
turbulence computations. In inverse cascades, it is the small scales 
which feed some cascadable quantity to the large scales, and any 
departure from proper dynamics at the small scales might be expected to 
have non-trivial implications for the evolution of the large scales.

The way inverse cascades have been studied numerically in the past, starting 
apparently with Lilly \cite{lilly}, is to write, on the right hand sides of 
the dynamical equations, “forcing terms” which are external to the fluid or 
magnetofluid equations, and which are there to inject excitations in some 
field or another. These terms are typically band-limited in Fourier space, so 
that only a narrow range of $k$-values, well above the energy-containing 
scales, are considered as externally excited or stirred. The excitations 
can be injected either into the mechanical or the magnetic part of the 
dynamics. In studies of the three-dimensional “dynamo”  problem, the 
injection is typically into the velocity field, and involves the 
conversion of mechanical helicity into magnetic helicity, which is then 
transferred back into the long-wavelength part of the spectrum 
(see, \eg, Refs. \cite{mazure}-\cite{montgomery}). 

In two dimensions, the helicities are identically zero, and the “anti-dynamo” 
theorem prohibits the generation of persistent magnetic excitations by 
mechanical means, so the band-limited  injections are magnetic and typically 
are considered to be the addition of mean square vector potential or magnetic 
flux, added randomly at the small scales 
(see, \eg, Refs. \cite{FM76}-\cite{HMM83}). 
Here, we write a random forcing function on 
the right hand side of Eq. (\ref{E16}), which may be described as follows.
We adopt a random forcing $f$ only in the induction equation for the vector 
potential, of the form
\begin{equation}
f(\bx,t) = \sum_{\bk} \hat f(\bk,t) e^{i \bk \cdot \bx} .
\end{equation}
The sum runs from $k_1=18$ to $k_2=22$. The amplitudes of all the coefficients 
$\hat f(\bk,t)$ in this ring are chosen equal, but the phases at each $k$ are 
changed randomly with a correlation time larger than the time step, but
smaller than the eddy turnover time. The phases are uniformly distributed 
between $-2\pi$ and $2\pi$. For the runs we will discuss in this 
Section, the correlation time was $\tau = 2 \times 10^{-1}$. 
One would expect that the relation of the “forcing band” of wave numbers to 
the reciprocal of $\alpha$ would be a sensitive one in the outcome of an 
inverse vector potential cascade computation. This proves to be the case, and 
only in the situation where the forcing band lies at lower wave numbers than 
the reciprocal of $\alpha$ are recognizable results achieved. Even there,  as 
will be seen in what follows, the agreement is less satisfactory than it has 
been for the selective decay and dynamic alignment tests.

Figures \ref{fig9}a,b show the time histories of a magnetically forced run 
that started from an otherwise empty spectrum, with a time step
$\Delta t = 2 \times 10^{-2}$, and $\eta = \nu = 10^{-3}$; the Taylor Reynolds
numbers at peak dissipation is for all runs $\sim 30$.
The upper curves in Fig. \ref{fig9}a are magnetic energies and the lower set 
are kinetic energies. The curves in Fig. \ref{fig9}b are 
mean square vector potentials as functions of time.
The solid lines are for a full MHD run ($\alpha = 0$) with $256^2$ grid 
points. The dashed lines are for $\alpha = 1/20$ and $256^2$ grid points. The 
dashed-dotted lines are for $\alpha = 1/30$ 
and $256^2$ grid points. The dashed-triple-dotted lines are for the mechanical 
$\alpha_k = 1/20$, but with the alpha parameter appearing in the induction 
equation $\alp_m$
set equal to zero (\ie the magnetic variables are unsmoothed).
The dotted lines are for $128^2$ grid points and both alphas $= 1/30$. 
The forcing functions are identical in all cases. Both kinetic and magnetic 
energies are similar in amplitudes. The biggest disparity will 
be noted in Fig. \ref{fig9}b, where the growth rates of the global 
mean-square vector potential differ significantly, resulting in ${\cal A}$ 
for the DNS run remaining about a factor of 2 larger than for any of 
the alpha approximations at the end of the runs.

\begin{figure}
\includegraphics[width=7.4cm]{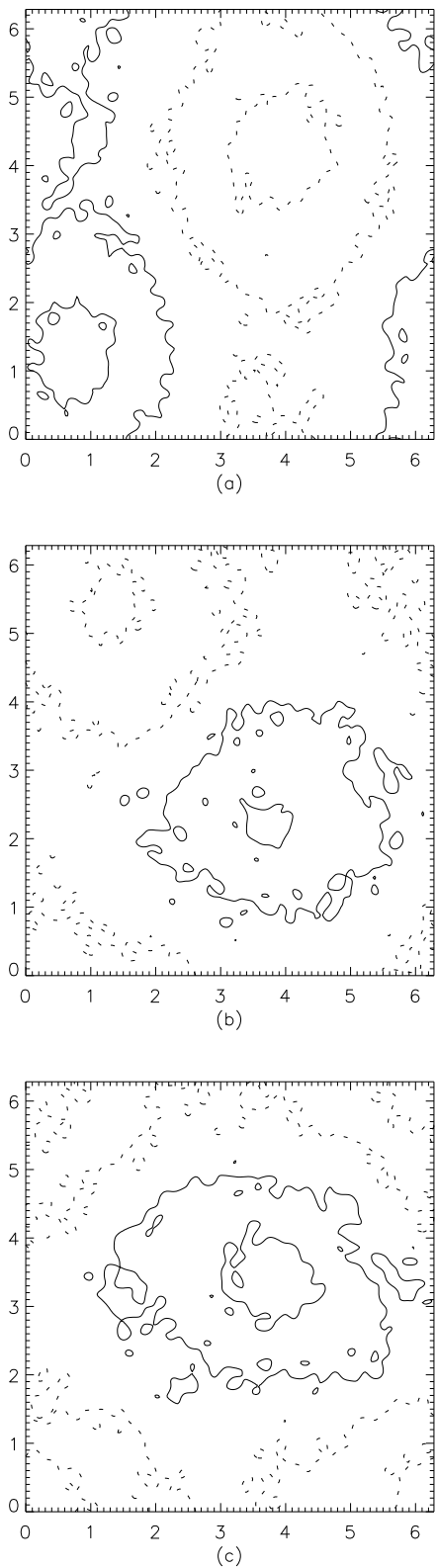}
\caption{Contour plots of the (smoothed) vector potential at $t=400$ for 
    inverse cascades (top: DNS; middle and bottom: runs 11 and 10 
    respectively). By that time, similar large-scale structures have 
    formed in all runs, but note that they have different locations in 
    the DNS and alpha runs.}
\label{fig12} \end{figure} 

Most of the discrepancy is accounted for by the values of the $k_{min}=1$ 
modes, the fundamentals, which are known from previous MHD inverse cascade 
computations to run away, at long times, from their nearest neighbors until 
limited by their own dissipation rates \cite{HMM83}.
Throughout the rest of the spectra below the 
forcing band, the disagreement is not so severe, as seen in Figs. \ref{fig10}a 
(vector potential spectra), \ref{fig10}b (magnetic energy spectra) and 
\ref{fig10}c (kinetic energy spectra). The resolutions, values of $\alpha$, 
and plotting conventions are the same as in Figs. \ref{fig9}. The magnetic 
potential is seen to follow a power law in agreement with the 
Kolmogorov-like estimation derived in Ref. 
\cite{P78}, \viz $\sim k^{-7/3}$; the ensuing spectrum for the 
magnetic energy is almost flat, $\sim k^{-1/3}$. The kinetic energy spectrum 
on the other hand follows approximately a $k^{1/3}$ law, also found in 
\cite{biskamp2d}; because the velocity
field is (partially) slaved to the magnetic field (kinetic energy is 
$\sim 20$\% of its magnetic counterpart), at longer times, Alfv\'en waves 
can put in rough equipartition the kinetic and magnetic modes, except in the 
forcing band, and it is expected that $E_k\sim E_m$ below the forcing band.
We also note that, at the onset of the inverse cascade process at early times, 
the spectra develop at large scales first a 
$k^5$ spectrum, followed in time by a $k^3$ spectrum (not shown), both for the 
full MHD run and for the alpha runs. These spectra are expected because of 
back-scatter (see also Ref. \cite{biskamp2d} for similar results).

Comparably wide disparities are visible in the spectral flux function 
$\Pi_A(k)$ for the squared vector potential, shown as functions 
of wavenumber $k$ in Fig. \ref{fig11} at time $t = 400$. 
Defining first the transfer of magnetic potential $T_A(k)$, in Fourier space, 
we have as usual for the DNS run:
\be
T_A(k)= \int \hat A^*_{\bk}\cdot {\cal F}({[\Psi, A])}_{\bk} d\theta_{\bk} 
\ + c.c.\ ,
\label{trans}\ee
with a similar definition for the alpha model; $c.c.$ means complex conjugate,
all fields in the above 
equations are taken to be smooth, the $*$ indicates complex conjugate,
and ${\cal F()}$ stands for taking a
Fourier transform of the whole Jacobian bracket. The flux is then 
defined as usual from the transfer as:
\be
\Pi_A(k) = \int_0^k dp\ T_A(p) \ .
\label{flux}\ee
The plotting conventions are the same 
as in Figs. \ref{fig9} and \ref{fig10}. None of the alpha-model attempts comes 
close to the true MHD flux function (solid line), a phenomenon exemplified by
the linear scale used here.

\begin{figure}
\includegraphics[width=9cm]{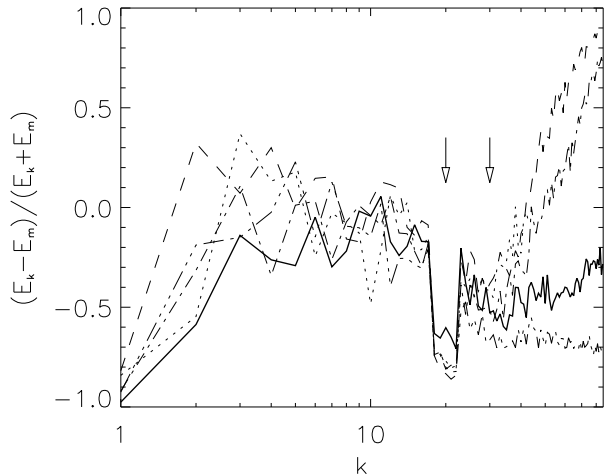}
\caption{Residual energy spectrum $E_k(k)-E_m(k)$ normalized by the 
    total energy $E_k(k)+E_m(k)$ at $t=400$, with plotting conventions 
    as in Fig. \ref{fig10}. The two vertical arrows correspond to the 
    two values of alpha, one of which is in the middle of the forcing 
    band (run 9, dashed line).  Note the strong dominance of kinetic 
    energy in the small scales for all runs except the DNS run (solid 
    line) and for the alpha run in which no smoothing occurs for the 
    magnetic field, \ie with $\alp_m=0$ (dash-triple dotted line).}
\label{fig13} \end{figure}

In Figs. \ref{fig12}, we display contour plots of curves of constant $A_{s_z}$
at several different times. The small-scale waviness of the large scale 
contours is due to the forcing near $k\sim 20$.
We can see that the details of 
the $A_{s_z}$-contours cease to be well reproduced by the alpha-model 
approximations, 
despite the agreements through most of the long-wavelength parts of the 
spectra. As in the earlier cases, the accurate reproduction by the alpha model 
seems to be confined to the long-wavelength spectral components, but does not 
include the detailed locations of individual features in configuration space.

\begin{figure}
\includegraphics[width=9cm]{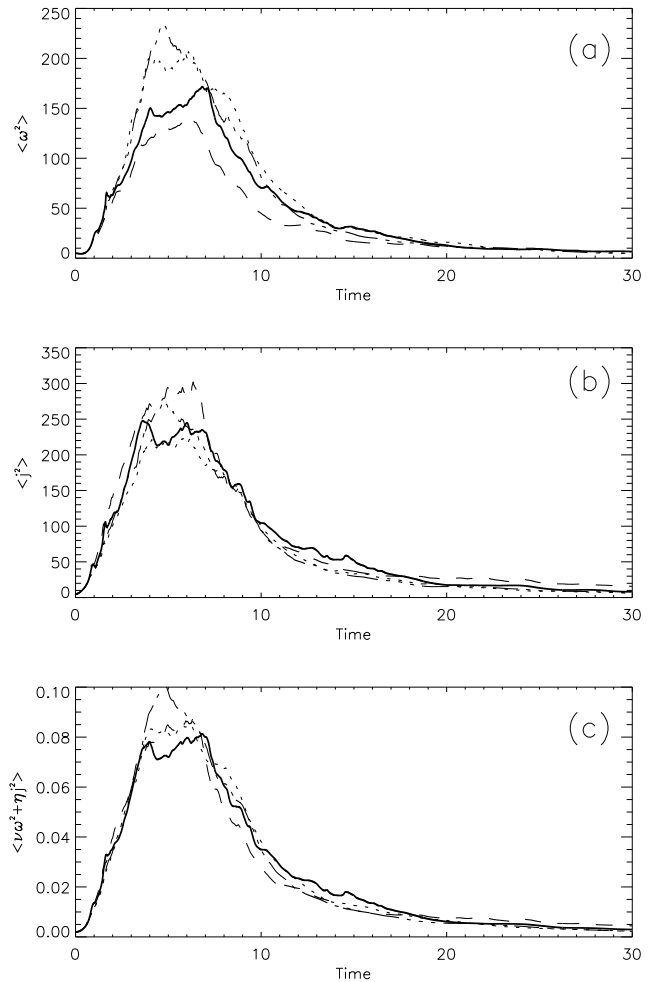}
\caption{Averaged (a) enstrophy, (b) square current and (c) total 
   dissipation for all runs for large-scale decay (runs 13-16, 
   see Table \ref{tabr}). Note that the maxima are reached at the 
   same time, but with an over estimation of gradients in the 
   alpha runs.}
\label{figdiss} \end{figure}

\begin{table}
$$\vbox{\offinterlineskip\halign{\tv#&\cc{#}&
\tv#&\cc{#}&\tv#&\cc{#}&\tv#&\cc{#}&\tv#&\cc{#}&\tv#&\cc{#}&
    \tv#&\cc{#}&\tv#\cr
\noalign{\hrule}
\traithorizontal
\tvi &run&&16$_m$&&14$_m$&&15$_m$&&16$_k$&&14$_k$&&15$_k$&\cr
\traithorizontal \traithorizontal
\tvi &$E_1^{D}$&&.15&&.08&&.15&&.17&&.12&&.15&\cr
\traithorizontal
\tvi &$E_2^{D}$&&.03&&.005&&.03&&.03&&.02&&.02&\cr
\traithorizontal \traithorizontal
\noalign{\hrule}}}$$
\caption{Errors (see eq. (\ref{er4})) for selective decay alpha runs 
    14--16, the subscripts $m$ and $k$ indicating respectively, the 
    error computed on the magnetic and kinetic energy spectra.} 
\label{tab1} \end{table}

\begin{figure}
\includegraphics[width=9cm]{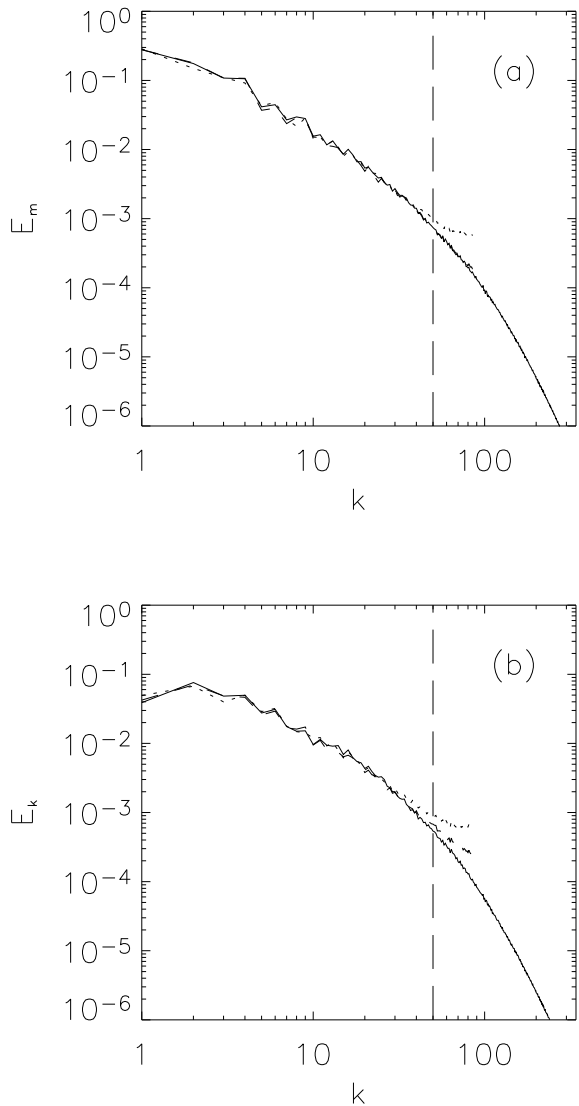}
\caption{Averaged (a) magnetic energy and (b) kinetic energy spectra 
   for the large-scale turbulence runs 13--16 (see Table \ref{tabr}) 
   between $t=3$ and $t=7$ (first and second peak of the enstrophy). 
   The wavenumber 
   corresponding to $\alpha$ is indicated by the vertical line, and the 
   under-resolved run is shown with a dotted line. At these early times, 
   spectra are in close agreement except at small scales.}
\label{figE1} \end{figure}

Finally, note that the inverse cascade phenomenon in two-dimensional 
Navier-Stokes turbulence has been studied using the alpha model in Ref.
\cite{nadiga} where it is argued that, in that case, the 
inverse cascade of energy to large scales is enhanced, quite substantially, 
when the alpha model is switched on; however, the fact that in Ref. 
\cite{nadiga} both a sink at long wavelengths and hyperviscosity
at short wavelengths are utilized may be at the source of this difference 
in behavior, compared to the case studied here. In that spirit, we can examine 
the kinetic energy defect in the small scales defined as
\be
R(k)= E_k(k)-E_m(k)\ ,
\label{defe} \ee
normalized by the total energy density $E_k(k)+E_m(k)$ at that 
wavenumber, with $E_k$ and $E_m$ the kinetic and magnetic energy spectra; 
as usual, when $\alp\not= 0$, the spectral density corresponding to the 
invariant (here, the total energy) will be taken. This energy imbalance 
(\ref{defe}), when integrated over small scales, is known to be
a source of the large-scale magnetic enhancement exemplified by the
inverse cascade of magnetic potential, through a mechanism akin 
to a negative eddy diffusivity (here, resistivity), as shown in 
Ref. \cite{P78} using a second-order 
closure of turbulence, or in \cite{BW83}-\cite{MH84} using an assumption of
scale separation. In other words, it represents non-linear non-local 
interactions in Fourier space, but it does not say anything about the local 
interactions themselves. 
In Figure \ref{fig13}, we plot $R(k)/[E_k(k)+E_m(k)]$ for several runs,
with as usual the solid line for the MHD run and the dash-triple-dot line for 
the alpha run without smoothing of the magnetic field. 
Note first that in all the runs, the fundamental mode is completely dominated 
by the
magnetic energy, and in all the alpha-runs except the one without smoothing 
of the magnetic field ($\alp_m=0$), the small scales are completely dominated 
by the velocity; this effect is enhanced by the normalization, but 
little energy resides in the smallest scales, beyond alpha, so that the inverse
cascade can still take place in alpha runs, albeit at a slower pace.
In other words, we see that in the framework of the alpha model,
this energy defect is modified from the usual MHD case, and more importantly
it even changes sign for the larger values of $\alp$ at high wavenumber, 
becoming positive and thus indicative of too high a dissipation of the 
magnetic field in the presence of alpha-smoothing of small scales, \ie with
$\alp_m =0$.
Even though alpha modeling is about the dynamical evolution of scales larger 
than alpha, the scales smaller than alpha nevertheless participate into the 
dynamics of the flow including at large scale and are responsible for the
discrepancy in the growth rate of squared magnetic potential we observe here
(see Fig \ref{fig9}b).
Thus, such an effect presumably
linked to the fact that the magnetic field is decaying with a scale
dependency that is leaning more heavily on the small scales (like an
effective $k^4$-type hyper-resistivity), may be at the source of the different 
behavior we observe between the different alpha-runs and the DNS run; if only
non-local transfer (in Fourier space) is active in the inverse cascade in the 
case of the alpha runs, this could be at the origin of the slower growth of 
${\cal A}$.
This point will require further investigation in the three-dimensional case,
in relation with the large-scale dynamo problem, since the growth rate
associated with the inverse cascade of magnetic helicity is the
relative (kinetic helicity minus magnetic current helicity) in the small 
scales \cite{strong}.
In the limit of very large $\alpha$, the dynamics become trivially 
simple; see Ilyin and Titi \cite{Ilyin03}.

\section{QUANTITATIVE ESTIMATES OF ERROR}

Two separate questions of accuracy are addressed in this Section. First,
we attempt quantitative comparisons of alpha model computations run on a
$256^2$ grid with the standard provided by a well-resolved MHD run at a
resolution of $1024^2$ for the same initial conditions and the same Reynolds
numbers.  Secondly, we compare the same alpha-model runs with DNS solutions 
on a $256^2$ grid in which no smoothing has been applied and
which are undoubtedly unresolved, in that the dissipation wavenumbers
(estimated on the basis of viscous and Ohmic enstrophy dissipation)
exceed the maximum values of $k$ retained in the computation; such runs
correspond to under-resolved computations. This latter
test is necessary if one is to argue that accuracy has been improved by
the alpha model in computations of comparable resolution.
Error estimates are provided first for freely-decaying turbulence
at early times, near the peaks of the dissipations (runs 13--16 of Table 
\ref{tabr}), and then for the
situations described in Secs. IV and V, where later times (and hence
greater accumulated errors) are involved. 

First we examine the case of freely
decaying turbulence at an early time, which could be considered to be
the early stages of a selective decay computation. 
The turbulence initially is confined to a band of wavenumbers
between $k_1=1$ and $k_2=3$.
In Figure \ref{figdiss}, we show the computed mean-square vorticity (a),
mean square current density (b), and mean square dissipation 
$\nu <\omega^2> + \eta <j^2>$ as functions of time. We show runs under
all four circumstances, starting from the same initial conditions:
fully-resolved $1024^2$ MHD, an alpha model run at $256^2$ with 
$\alpha = 1/50$ for both fields, an alpha model run with an unsmoothed 
magnetic field ($\alp_m=0$) and $\alpha_k=1/50$ for the velocity field 
at $256^2$, and an under-resolved but unsmoothed DNS run at $256^2$;
we see that time scales of growth of field gradients, and the amplitudes of 
such gradients are comparable but not identical.

\begin{widetext} \begin{center} \begin{table}
$$\vbox{\offinterlineskip\halign{\tv#&\cc{#}&
\tv#&\cc{#}&\tv#&\cc{#}&\tv#&\cc{#}&\tv#&\cc{#}&\tv#&\cc{#}&\tv#&\cc{#}
    &\tv#&\cc{#}&\tv#&\cc{#}&\tv#\cr
\noalign{\hrule}
\traithorizontal
\traithorizontal
\tvi&run&&$E_m$: 12&&$E_m$: 9&&$E_m$: 10&&$E_m$: 11&&$E_k$: 12&&$E_k$: 
    9&&$E_k$: 10&&$E_k$: 11&\cr
\traithorizontal
\traithorizontal
\tvi&$E_1^{I}$&&.30&&.49&&.31&&.28&&.28&&.33&&.26&&.26&\cr
\traithorizontal
\tvi&$E_2^{I}$&&.13&&.41&&.14&&.11&&.09&&.13&&.08&&.09&\cr
\traithorizontal
\traithorizontal
\noalign{\hrule}}}$$ 
\caption{Errors as defined in eq. \ref{er4} for the inverse cascade alpha 
  runs 9--12 (see Table \ref{tabr}); $E_m$ and $E_k$ indicate respectively 
  the magnetic and energy spectra. Note the larger errors for run 9 
  corresponding to the alpha wavenumber ($k_{\alpha}=20$) embedded in the 
  forcing band.}
\label{tab2}  \end{table} \end{center} \end{widetext}

Note that for the fully resolved DNS run, $k_{max}=341$
to be contrasted with the dissipation wavenumber based on the magnetic 
variables (computed using a Kolmogorov 
spectrum) of $k_{diss}\sim 280$ at the peak of dissipation \ie for $t\sim 7$
(and $k_{diss}\sim 250$
when based on the velocity); this shows that the run is well resolved at all 
times, and indeed no ``bottle-neck'' appears in the compensated spectra, where 
by bottle-neck is meant an
accumulation of energy in the small scales, seen as a bump in the 
compensated spectra for $k\sim k_{max}$; this phenomenon has sometimes
appeared in the presence of under-resolved computations, or when using 
hyper-viscous or hyper-resistive dissipative operators
\cite{biskamp2d}. 
At the same time $t\sim 7$, the Taylor Reynolds number (see equation 
(\ref{taylor_r})) is $R_{\lambda}\sim 1150$, with a Taylor wavenumber
$k_{\lambda}\sim 20$ (with $k_{\lambda}\sim 17$ when based solely on the 
current density); in this case, only run 14 (with both alphas non zero)
has a slightly higher $\Rla$ at that time, whereas runs 15 (with $\alp_m =0$)
and run 16 (under-resolved DNS) have slightly lower $\Rla$.

\begin{figure}
\includegraphics[width=9cm]{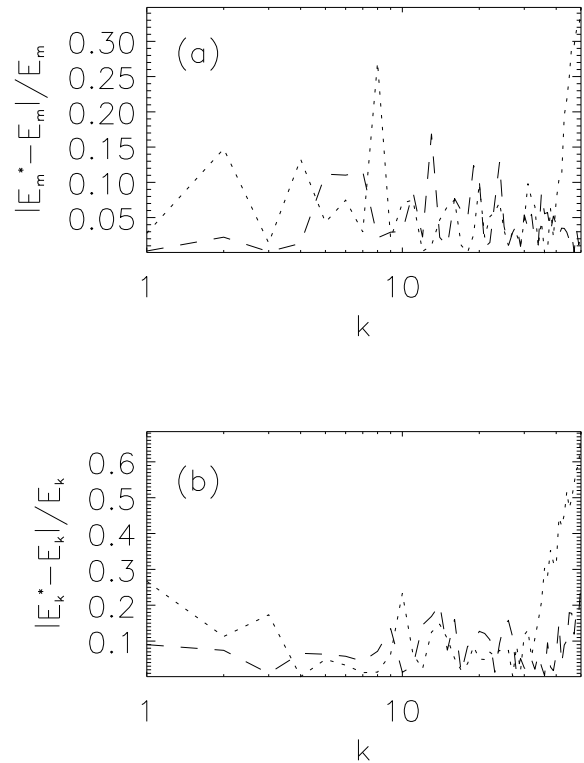}
\caption{Normalized errors, shown up to $\alpha^{-1}$, in time-averaged 
   (a) magnetic 
   and (b) kinetic energy spectra for the large-scale turbulence runs 13--16.
   Larger discrepancies occur both at small 
   and at large $k$, in particular for the unresolved run (dotted line).}
\label{figE2} \end{figure}

Figure \ref{figE1} shows the energy spectra, for the freely decaying run, 
averaged over the period of time when the dissipation remains 
quasi-constant ($t=3$ to $t=7$), and Fig. \ref{figE2} displays the 
relative error spectrum, defined as 
\be
e_r(k)=\frac{E^*(k)-E(k)}{E(k)} \ .
\label{err}\ee
Here, $E(k)$ is the well-resolved energy spectrum (either kinetic or 
magnetic), summed over all $k$-vectors with the same magnitude, regarded 
as the ``truth,'' and $E^*(k)$ is the energy computed from any one of the 
stated three approximations to it. 

\begin{figure}
\includegraphics[width=9cm]{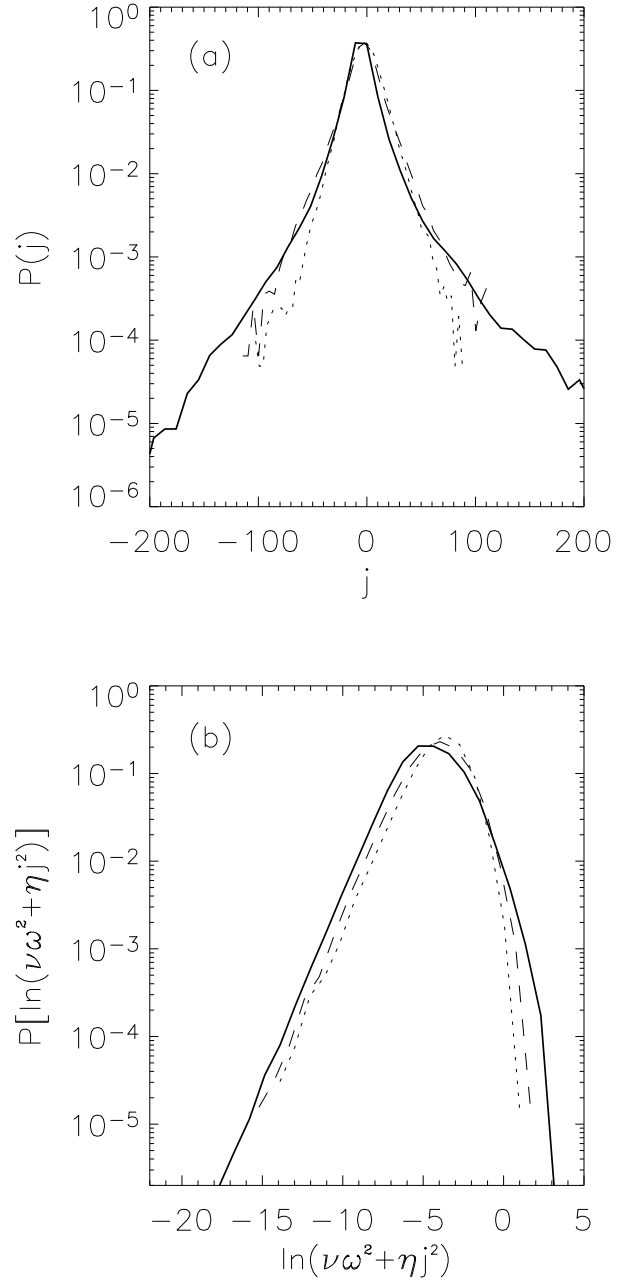}
\caption{Normalized pdfs of (a) the current density, and (b) 
   $\ln(\nu \omega^2 + \eta j^2)$, 
   for the large-scale turbulence runs (see Table \ref{tabr}).
   At this time, small-scale structures are less developed in the alpha runs
   than in the DNS run (solid line).}
\label{figE3} \end{figure}

Both figures show that the under-resolved flow displays systematically a 
greater error than the alpha model run, at large scales ($k=1$ to $k=3$) 
as well as at small scales near the cut-off, $1/\alpha$. Finally, in 
Fig. \ref{figE3}, are plotted the pdfs of (a) the current and (b) the 
spatial density of the dissipation of energy, the
former in lin-log scales, and the latter in log-log scales. The same line
conventions as in Fig. \ref{figE1} have been adopted. The alpha model clearly
reproduces better the gradients, at a given resolution, than the 
under-resolved flows.

Two possible global quantifications for measuring the errors are E1 and E2, 
defined by the relations:
\be
E_1= \frac{\Sigma_{1}^{1/\alp}|E^*(k)-E(k)|}
{\Sigma_{1}^{1/\alp}E(k)}, \ 
E_2= \frac{\Sigma_{1}^{1/\alp}(E^*(k)-E(k))^2}
{\Sigma_{1}^{1/\alp}E^2(k)} \ .
\label{er4}\ee
Thus a small value of $E_1$ or $E_2$ will indicate
a closeness on the part of the MHD approximations (alpha-modeled
or unresolved) to the full MHD DNS results.

$E_1$ and $E_2$ are exhibited as Table \ref{tab1} for the freely decaying 
runs just described (and indicated by the superscript ``D''). The 
conventions used in the Table are that the number of the run is followed 
by a subscript, where $m$ stands for magnetic spectra whose error is 
being assessed, and $k$ for kinetic ones.
Time averages have been performed from $t=3$ to $t=7$, the vicinity of 
the main peaks in the dissipation rates. This is thought to be the time 
when the turbulence is of its most broad-band character, when the alpha 
model would be having its strongest impact.

The errors so defined are smaller for these early times for this
freely decaying situation than they are for the situations studied in
Sections III, IV, and V which examine late-time evolutions.
We note that the lowest error (by a factor of $\sim 2$ 
compared to the unresolved run) occurs for the alpha run with both alphas 
equal (run 14, see Table \ref{tabr}). 
Note also that Fig. \ref{figE2} shows errors at each $k$, whereas the errors
in Table \ref{tab1} are normalized by the total energy
(truncated at $1/\alp$) and hence are smaller.

In Table \ref{tab2}, we show the errors $E_1$ and $E_2$
for the inverse cascade situation of Sec. V
and as indicated by the superscript I; the runs are the same as those
in Sec. V, and given by their number (see Table \ref{tabr}), with $E_m$ and
$E_k$ standing as usual for magnetic and kinetic energy spectra.
These errors are
computed at late times, when the true solution may be expected to have drifted
further from the $1024^2$ run and hence lead to larger errors than in 
the freely decaying runs of Table \ref{tab1}.
Moreover, as expected, when the alpha cut-off is too close to the forcing band,
the errors are larger, both for the kinetic and the magnetic spectra.

Finally, in Table \ref{tab3}, the $E_1$ and $E_2$ errors are shown for the 
dynamic alignment runs 6 and 7 of Sec. IV, with a superscript ``DA'' and a 
supplementary index ``e'' or ``l'', in order to indicate early or late times
in the run; specifically, early signifies that the average is taken for 
for $5\le t \le 10$, and late for $55\le t \le 60$.
Again, the type of spectrum for which the error is displayed (either $E_m$ or
$E_k$) is given before the run number. 
The low-resolution computation (run 7) does not have significantly 
higher errors at early times, but errors accumulate at later times, more
so for the lower resolution computation (run 7). 
The reason for time averaging the magnitude of the errors is that, when 
plotted as functions of time, the error curves cross each other repeatedly. 
Time intervals can be found when either one is smaller than another. Time 
averaging, over intervals long enough to contain many of these crossings 
but short compared to the duration of the runs, has seemed to provide 
the most objective number for addressing which error is ``typically'' 
smaller.

\section{VERY HIGH REYNOLDS NUMBERS}

The eventual utility of the alpha model if it can be justified will be that it 
will permit explorations of Reynolds number regimes that are far above those 
that can be obtained from direct numerical solutions of the MHD equations. 
It may be noted that extimates have been given for the 
number of degrees of freedom of the Navier-Stokes alpha model 
\cite{Foias02}; see also \cite{Foias2001}. 
Whereas most of this paper has been devoted to regimes in which direct MHD 
solutions can be compared to alpha model solutions, we have thought it 
interesting to show one alpha model computation that goes beyond what can 
be contemplated from unsmoothed solutions presently. We do this without any 
definitive claims for accuracy, but just as a suggestion of what the alpha 
model might provide in the way of future predictions. A detailed analysis of
high Reynolds number runs at higher resolutions than what is performed here and
using the alpha model will be presented elsewhere.

\begin{figure}
\includegraphics[width=9cm]{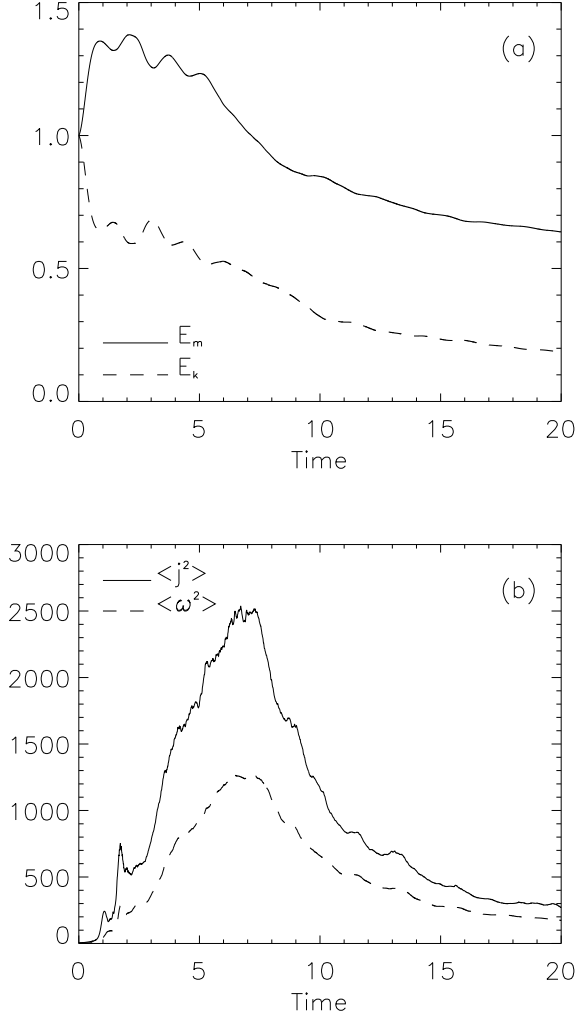}
\caption{(a) Magnetic and kinetic energy, and (b) enstrophy and square 
   current as a function of time for run 17 (see Table \ref{tabr}) with 
   $2048^2$ grid points and $\alpha^{-1}=300$. Note the several peaks 
   in the enstrophy and the mean square current density, as well as 
   the excess in magnetic excitation, both at large scale and at small 
   scale.}
\label{figreyt}
\end{figure}

We display results for a run with $\eta = \nu = 2 \times 10^{-5}$, a time step 
of $\Delta t=2.5 \times 10^{-4}$, and $2048^2$ grid points. The value of 
$\alp^{-1}$ is chosen to be $300\sim k_{max}/2$, with $k_{max}\sim 682$ for 
this run.
The initial (equal) kinetic and magnetic energies are loaded with random 
phases into the ring from $k_1=1$ to $k_2=3$ in Fourier space, and the \rms 
values of $\bv$ and $\bB$ are unity. Computed at $t=7$, \ie close to the 
maximum of dissipation (see Fig. \ref{figreyt}b), the Taylor Reynolds number 
$R_{\lambda}\sim 5200$. This is roughly comparable with what can be 
accomplished with a DNS on a grid of more than $10^8$ points.

Fig. \ref{figreyt}a shows the evolution of the total kinetic energy 
(dashed line) and total magnetic energy (solid line),
referred as usual to unit volume, 
as functions of time. Fig. \ref{figreyt}b shows the evolution of the 
mean square vorticity (dashed line) and mean square current density 
(solid line) as functions of time. The qualitative behavior of all 
quantities in Figs. \ref{figreyt} will 
be seen as not significantly different from that observed at lower 
resolutions, although oscillations in the kinetic and magentic energies are
persistent until $t\sim 6$. But the idea has been to extend the inertial range 
as much as possible. It will be noted that the magnetic and kinetic energies 
are far from equipartitioned, with magnetic energy in excess by a factor 
$\sim 3$ by the end of the run. The peak value of mean dissipation at $t\sim 7$
($\bar \epsilon\sim 0.074$)
is comparable with that at the lower Reynolds number (runs 13-16), confirming 
previous results (see \eg Fig. 7 in Ref. \cite{politano89}) of lack of
dependency of $\bar \epsilon$ with Reynolds number, at least at a magnetic 
Prandtl number of unity.

\begin{table}
$$\vbox{\offinterlineskip\halign{\tv#&\cc{#}&
\tv#&\cc{#}&\tv#&\cc{#}&\tv#&\cc{#}&\tv#&\cc{#}&\tv#\cr
\noalign{\hrule}
\traithorizontal \traithorizontal
\tvi &runs  && $E_m$:\ 7  && $E_m$:\ 6  && $E_k$:\ 7 && $E_k$:\ 6  &\cr
\traithorizontal \traithorizontal
\tvi&$E_1^{DA_e}$&& .17 && .19&& .23&& .22  &\cr
\traithorizontal
\tvi&$E_2^{DA_e}$&& .03 && .04&& .08&& .06 &\cr
\traithorizontal \traithorizontal
\tvi&$E_1^{DA_l}$&& .33 && .27 && .43 && .44 &\cr
\traithorizontal
\tvi&$E_2^{DA_l}$&& .13 && .07 && .33 && .23 &\cr
\traithorizontal \traithorizontal
\noalign{\hrule}}}$$
\caption{Errors on the magnetic $E_m$ and kinetic $E_k$ energy spectra 
   for dynamic alignment runs (see Table \ref{tabr}) for early (e) 
   and late (l) phases of the evolution. As expected, errors are larger at 
   later times, and are larger for the low resolution computation (run 7).}
\label{tab3} \end{table}

\begin{figure}
\includegraphics[width=9cm]{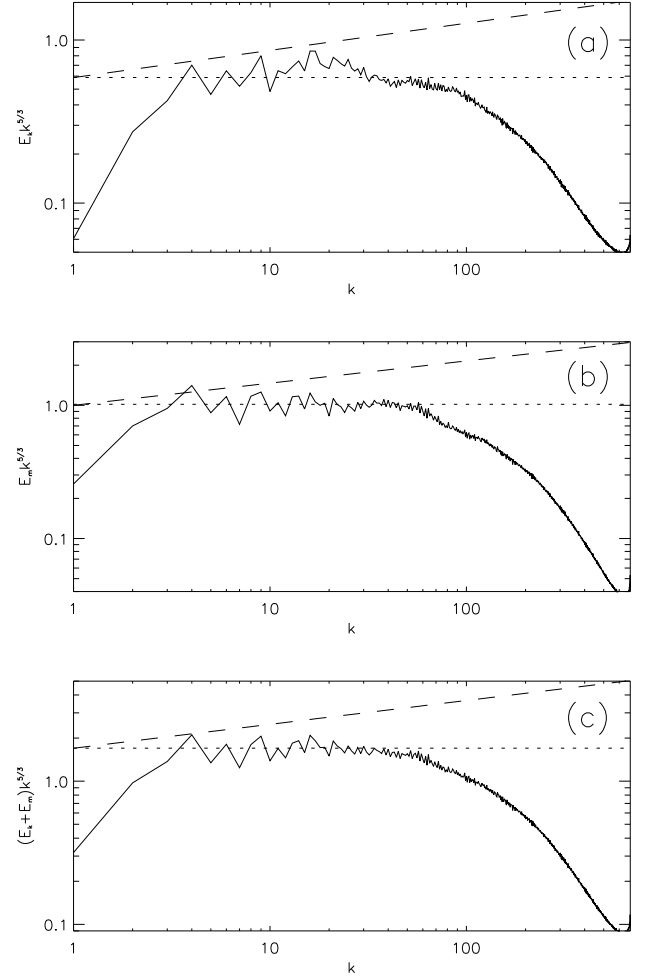}
\caption{(a) Kinetic energy spectrum, (b) magnetic energy spectrum, and 
   (c) total energy spectrum compensated with $k^{5/3}$, for run 17 (see 
   Table \ref{tabr}). Temporal average is performed between $t=2$ and 
   $t=6$. The horizontal dotted line corresponds to a $k^{-5/3}$ spectrum, 
   and the dashed line to a $k^{-3/2}$ law. A clear inertial range extends 
   for more than one decade in wavenumber.}
\label{figreys} \end{figure}

Compensated energy spectra are displayed in Figs. \ref{figreys}. 
Fig. \ref{figreys}a is the kinetic energy spectrum, multiplied by $k^{5/3}$, 
and averaged between $t=2$ and $t=6$, \ie the times at which the energies are
oscillating with little dissipation yet. Fig \ref{figreys}b shows the magnetic 
energy spectrum averaged over the same time interval and similarly multiplied 
by $k^{5/3}$. Fig. \ref{figreys}c shows the total 
energy spectrum, similarly compensated and time averaged over the same 
time interval.
In all three figures, the horizontal dotted line has zero slope, and so would 
coincide with a $k^{-5/3}$ Kolmogorov inertial range spectrum so compensated. 
In all three 
figures, the dashed line has slope $1/6$, and so would be tangent to 
a $k^{-3/2}$ spectrum (as proposed by Iroshnikov and Kraichnan \cite{iro,rhk})
which had been multiplied by $k^{5/3}$.  It would appear from these figures 
that $k^{-5/3}$ would fit the computed spectra significantly better 
than $k^{-3/2}$ would.
We do not attach any finality or conclusiveness to this observation, 
because intermittency is known to steepen energy spectra obtained from
dimensional analysis, and it is known that high-order structure functions 
computed for statistically steady 2D-MHD flows display a behavior that differs 
from that of turbulent neutral (3D) fluids \cite{poli1}.
The total energy spectrum computed with the alpha model
agrees with other findings (see \eg \cite{poli1,poli2} \cite{biskamp2d}) at 
lower Taylor Reynolds numbers. In that light, we conclude that the alpha
model does not alter previously known results, suggesting that intermittency is
worth further investigation in the context of the alpha model, both in two and 
three dimensions. 
In the same spirit, we note the absence of any ``bottle-neck'' in the spectra, 
even at these high Reynolds numbers, in contrast to what is found in 
\cite{biskamp2d}.

Finally, it is worth noting that, computed at the
maximum of dissipation at $t\sim 7$, the Taylor wavenumber in the alpha run is 
$k_{\lambda}\sim 55$, \ie well below the alpha wavenumber of $300$, whereas
the dissipation wavenumber, based on the alpha-enstrophy $<\omega \omega_s>$ 
and the square current $<j^2>$, are respectively $1300$ and $1500$,
\ie well beyond the largest resolved wavenumber ($k_{max}=682$) for this
computation; however, even when plotting the dissipation spectrum 
$k^2[E_k(k)+E_m(k)]$, no bottle-neck is observed (not shown).

\section{SUMMARY AND DISCUSSION}

The intent of this article has been an empirical study of the extent to which 
the alpha model, or Lagrangian-averaged model, equations predict the 
results of computed incompressible 2D MHD behavior in which no further 
modeling or approximations are made. Our primary motivation has been to 
acquire confidence in the alpha model in hopes that it can be used for 
problems with such high Reynolds-like numbers that they cannot be computed in 
the framework of the primitive MHD equations. A useful starting point has 
seemed to be to work on classical problems in rectangular periodic 
conditions where some information about solutions has accumulated over the 
last thirty years. These problems include those often grouped under the terms 
selective decay, dynamic alignment, direct and inverse cascades, and 
tabulating frequency distributions or pdfs for fluctuating field quantities. 
We have compared the results of direct numerical solutions for these problems 
with the solutions of alpha-model equations for the same initial and boundary 
conditions, both for freely decaying and for forced turbulence. 
In order to do so at sufficiently high Reynolds numbers, we have kept our
comparisons to the two-dimensional geometry for which reasonable resolutions 
can be obtained without overly taxing available computer resources.

The principal success we have to report is in the satisfactory reproduction 
of the long-wavelength spectral behavior for magnetic and velocity field 
evolution, in nearly all cases. Also, characteristic length scales of the 
flow, as the Taylor scale, are well reproduced by the alpha modeled 
simulations. The only exception is that some discrepancies remain for 
the longest-wavelength behavior in the case of driven cascades of 
mean-square magnetic vector potential at late times. The model has 
not proved accurate in the detailed reproduction of specific spatial 
features of the turbulence at earlier times. Nor does the alpha model 
reproduce accurately the pdfs of intermittent fluctuations of large 
amplitude, and likely because of the deliberate suppression of small 
spatial scales, is not expected to.

It should be remarked that there is the more difficult problem of obtaining 
a clear physical understanding of the nature of the alpha approximation itself.
It was originally arrived at by mathematics of considerable sophistication and 
complexity,
which left intuitive gaps in just what was being assumed. This is a very 
different perspective than the one used in Ref. \cite{MP02} or Section II 
of this paper, where the recipe of smoothing the $\bv$ and $\bB$ fields but 
not their sources, and then neglecting the fluctuations in those fields 
about their averages was invoked. Neither prescription seems clear 
enough to us at 
a physical level to argue for it strenuously on any basis other than its 
satisfactory consequences. Nor is it clear why two such apparently different 
procedures should end at the same place, despite the happy fact that they 
seem to do so. Clarifying the conceptual foundations of such modeling
at an intuitive physical level is a serious but stimulating challenge.

In summary, our judgment of the alpha model is that it reproduces 
satisfactorily the time development given by well-resolved DNS computations 
for spectra up to about $k \sim 1/\alpha$ and does well enough at 
reproducing the probability distribution functions of fluctuations. What 
it does not do satisfactorily is to reproduce the locations, trajectories, 
and shapes of structures in configuration space.

We see several directions in which to extend this work. For 
example, there is the natural one of three-dimensional computations, still in 
rectangular periodic boundary conditions, where such problems as the inverse 
cascade of magnetic helicity  (an inherent part of the large-scale “dynamo” 
problem), the spectral anisotropy induced by the presence of a dc magnetic 
field \cite{more1,more2}, and the small magnetic Prandtl regime
remain to be investigated 
(see for recent studies in the latter case \eg \cite{alex04,ppp04}).
Finally, the questions of material boundaries with non-ideal boundary 
conditions and departure from rectangular to spherical or cylindrical symmetry 
seem necessary as well. The only work published so far involving material 
boundaries and the alpha model seems to be that of Chen et al. 
\cite{other2,CFHOTW99}. 
It is our intent to move in these directions in the near future.

\begin{acknowledgments}
We thank Darryl Holm for helpful and stimulating discussions concerning
the alpha model, both for Navier-Stokes and MHD, Jean-Fran\c cois Pinton and 
Duane Rosenberg for a careful reading of the manuscript, and Henry Tufo for 
letting us compute on the NSF-ARI cluster at the University of Colorado.
The NSF grants ATM-0327533 at Dartmouth College and CMG-0327888 at NCAR
supported this work in part and are gratefully acknowledged.
Computer time was provided by equipment purchased under NSF ARI Grant
CDA-9601817 at the University of Colorado, and under NSF sponsorship of the 
National Center for Atmospheric Research.
\end{acknowledgments}

\end{document}